\documentclass[twocolumn,showpacs,preprintnumbers,amsmath,amssymb]{revtex4}
\usepackage{graphicx}
\usepackage{dcolumn}
\begin{document}

\title{$D=4$ supergravity dynamically coupled to
 superstring in a superfield Lagrangian approach}

\author{
Igor A. Bandos$^{\ast,\dagger}$ and Jos\'e M.
Isidro$^{\star}$ }
\affiliation{$^{\ast}$Departamento 
de F\'{\i}sica Te\'orica and IFIC (CSIC--UVEG), Universidad de Valencia,
 46100-Burjassot (Valencia), Spain. E-mail: bandos@ific.uv.es}
\affiliation{$^{\star}$ Instituto de F\'{\i}sica Corpuscular 
(CSIC--UVEG), Apartado de Correos 22085, Valencia 46071, Spain. 
E-mail: jmisidro@ific.uv.es}
\affiliation{$^{\dagger}$Institute for Theoretical Physics, NSC KIPT,
UA61108, Kharkov, Ukraine. E-mail: bandos@kipt.kharkov.ua }

\date{August 14, 2003, {\bf hep-th/0308102}}

\def\theequation{\arabic{equation}}
\begin{abstract}
We elaborate a full superfield description of the interacting system of
dynamical  $D=4$, $N=1$ supergravity and dynamical superstring.
As far as minimal formulation of the simple supergravity  is used,
such a system
should contain as well the tensor (real linear)
multiplet which
describes the dilaton and the two--superform gauge field whose pull--back
provides the Wess--Zumino term for the superstring.
The superfield action is given by the sum of the
Wess--Zumino action for $D=4$, $N=1$
superfield supergravity, the superfield action for
the tensor multiplet in curved superspace {\sl 
and the Green--Schwarz superstring
action}. The latter includes the coupling to the tensor multiplet
both in the Nambu--Goto and in the Wess--Zumino terms.
We derive superfield equations of motion including, besides the
superfield supergravity equations with the source, the source-full
superfield equations for the linear multiplet. The superstring equations
keep the same form as for the superstring in
supergravity and 2--superform background.
The analysis of gauge symmetries shows that the superfield description
of the interacting system is {\it gauge equivalent} to the dynamical system
described by the sum of the spacetime, {\it component} action for
supergravity interacting with tensor multiplet and of the
{\it purely bosonic} string action.

\end{abstract}

\maketitle

\bigskip

\section*{Introduction}

Recently, there has been a new interest 
in the superfield description of supergravity
\cite{11SG,L11,PVW,HSS}. It is motivated,
in particular,  by the search for
a superfield formulation of 10--dimensional supergravity incorporating
the superstring corrections 
(see \cite{Gates86,Italy88,Mario} 
for early studies and \cite{Gates2002,Italy92} 
for discussions).   
Supergravity was known to appear at the
point--like limit of the supersting
which corresponds to 
$\alpha^\prime \rightarrow 0$ limit, {\it i.e.} to zeroth order in the
decomposition in the Regge slope parameter $\alpha^\prime$.
Already at the first order in  $\alpha^\prime$, the string corrections
modify the supergravity equations of motion. On the other hand,
the known superfield formulations  of  $D=10$ supergravity provide
its  on--shell description; it is given by the {\sl
on--shell} constraints on superspace torsion, which imply the dynamical
equations for the physical fields. There are just the equations
which correspond to the $\alpha^\prime \rightarrow 0$ limit of superstring.
Thus the incorporation of the $\alpha^\prime$ corrections requires the
modification of the standard superspace constraints:
a search for a possibility to replace the on-shell constraints by a set of 
off-shell constraints, or, at least, by a set of  'on any shell'
constraints \cite{11SG} including some parameters which specify the
right--hand side of the supergravity equations and which can be chosen to
describe the superstring corrections to such equations.

Basically the same problem appears when one searches for a superfield
description of a superbrane interacting with higher dimensional supergravity.
The superbrane is defined as a brane moving in superspace. It is well known
\cite{strKsg,BST87} that the requirement of a smooth flat superspace
limit for the superbrane in curved superspace
(which implies that the superbrane action in a curved superspace background
should possess the same number of gauge symmetries,
including fermionic $\kappa$--symmetries \cite{ALS})
results in the standard on--shell supergravity constraints.
However, as was said above, such constraints imply the `free'
supergravity  equations of motion without any superbrane source.
On the other hand, as is clear from the purely bosonic limit
(gravity interacting with a bosonic brane), the brane should provide a source
in the Einstein equation
\footnote{Note that a similar problem appeared for the heterotic superstring
\cite{Heterotic} in $D=10$, $N=1$ supergravity 
and $E_8\otimes E_8$ (or $SO(32)$)
super--Yang--Mills (SUGRA-SYM) 
background \cite{Mario}.
Namely, the requirement of
$\kappa$--symmetry of the classical heterotic superstring model
results in constraints which describe decoupled
SUGRA and SYM systems, while the Green--Schwarz anomaly
cancellation mechanism required their nontrivial interaction.
This problem had motivated the study  \cite{Mario} of
(one loop) quantum anomalies in the $\kappa$--symmetry
transformations. As was shown in  \cite{Mario}, such anomalies occur, but
may be absorbed by consistent quantum corrections to the classical
(tree--level) expressions for superspace
torsion and curvature. The new (one--loop)  torsion constraints
lead to the desired coupled equations for the SUGRA--SYM system.}.
So one could get the (mistaken) impression
that supersymmetry forbids interaction with an extended object, at least
at the classical level. Certainly this is not the case.
The resolution of such a paradox and a search for a  consistent
(quasi)classical description of the {\sl supergravity---superbrane} 
interacting systems
is of interest in its own, as well as in relation with its possible
applications to the study of quantum gauge theories in the language of
classical supergravity--superbrane models along the line of the $AdS/CFT$
correspondence \cite{AdS,YM,AdS1}.

The study of the complete Lagrangian superfield description of the
supergravity--superbrane interacting system, when it is possible,
{\it e.g.} in $D=4$, $N=1$ curved superspace,
might provide new insights into the search for a modification of
higher--dimensional
($D=10$, $11$) supergravity constraints in such a way that they would
produce dynamical equations with sources,
including singular sources from superbranes and nonsingular sources
describing the stringy corrections.

Such a study has been carried out in \cite{BAIL2} for the interacting system of
dynamical $D=4$, $N=1$ supergravity and a massless superparticle source.
Here we elaborate the superfield description of the next more complex
system which includes, besides dynamical $D=4$, $N=1$ supergravity,
the dynamical superstring. It has some specific features in comparison
with the system already studied in  \cite{BAIL2}. First,
the source is provided by a supersymmetric  {\sl extended} object.
Second, as far as the minimal formulation of $D=4$, $N=1$ supergravity
is considered, one finds (see \cite{G88}) 
that the Wess--Zumino term of the superstring describes a
coupling to an additional,  dynamical tensor (or real linear)
multiplet \cite{S79,1001} 
which can be used to formulate the nontrivial
2-superform gauge theory in superspace \cite{G80}. Moreover, 
superstring $\kappa$--symmetry requires the identification of the
tensor multiplet with a dilaton superfield and its coupling to
the kinetic, Nambu--Goto term of the superstring action.
Thus the superfield action for the interacting system which we will study
in this paper  includes, in addition to the Wess--Zumino action for
supergravity \cite{WZ78} and the
$D=4$, $N=1$ Green--Schwarz superstring action \cite{GS84},
also a superfield  action \cite{dWR} for the tensor
multiplet \cite{S79} in a curved superspace of 
minimal supergravity \footnote{
The sum of the 
superfield minimal supergravity action and a superfield  
tensor multiplet action was motivated by being a low energy limit of a 
$D=4$ $N=1$ compactification of the heterotic string 
\cite{S88}, \cite{G88}. 
Nevertheless,  this limit, as well as its $N>1$ generalizations,  
which was intensive studied in eighties and nineties   
\cite{4Dlimit,S95,G96,BS96}, is not a subject of the present paper.
We rather consider the $D=4$ supergravity--superstring interacting system as 
a relatively simple model for a more complicated 
$D=10, 11$ supergravity--superbrane systems. 
An interesting alternative possibility is to consider 
the new minimal formulation of superfield supergravity \cite{SW81},
where the supergravity {\sl auxiliary} fields are provided by 
a tensor multiplet, interacting with the Green--Schwarz superstring. 
This, however, is beyond the score of the present paper.}.
It has the form (see the main text for the notations)
\begin{eqnarray}\label{0Sint}
& S & = \int d^8 Z \, sdet(E_M{}^A)  \left(1 + s
{\Phi \over 2}
e^{{{\Phi}\over 2}}\right) +
{1\over 2\pi \alpha^\prime}  S_{sstr} \, , \; 
\\ \label{0GSac}
& S&\!\!\!\!{}_{sstr}  =
{1 \over 2} \int_{W^2} 
d^2 \xi \, e^{{\hat{\Phi}\over 2}}\, 
\sqrt{|det(\hat{E}_m^a \hat{E}_n^b \eta_{ab}|} 
-\int_{W^2}  \hat{B}_2 \; , \qquad
\end{eqnarray}
where $s$ is the coupling constant for the tensor multiplet,
${1\over 2\pi \alpha^\prime}$
is the superstring tension, which we set equal to unity in the main text of the
present article, $\xi^m=(\tau, \sigma)$ are local 
coordinates of the string worldsheet ${\cal W}^2$, 
$\hat{E}_m^a:= \partial_m \hat{Z}^M (\xi) 
E_M{}^A(\hat{Z}(\xi))$, the supervielbein $E_M{}^A(Z)$ and 2--superform
$B_2:={1\over 2}dZ^M\wedge dZ^N B_{NM}(Z)$
are subject to the constraints given in Sec. I, and, finally,
the superfield $e^{{{\Phi}\over 2}}$ satisfies the defining constraints
of a tensor multiplet in curved superspace,
\begin{eqnarray}
\label{0Phi=L}
({\cal D}{\cal D}-\bar{R}) e^{{{\Phi}\over 2}} =0 \; , \quad
(\bar{{\cal D}}\bar{{\cal D}}-R) e^{{{\Phi}\over 2}} =0 \; . \quad
\end{eqnarray}

The main result of this paper is
the complete set of superfield equations of motion for the
interacting dynamical  system (\ref{0Sint}),
 including the superfield supergravity equations and
the dynamical equations for the tensor multiplet
(\ref{0Phi=L}) with the superstring source  
\footnote{
Note that superfield equations of motion for the field theoretical part 
of our interacting system, 
{\it i.e.} following from the action
$S= \int d^8 Z \, sdet(E_M{}^A)  \Big(1 + s
{\Phi \over 2}
e^{{{\Phi}\over 2}}\Big) $, but without the $S_{sstr}$ term, 
were considered in \cite{S95,BS96}.}.  
This can provide some insight in a search for source-full
superfield equations for more complicated
interacting systems in $D=10,11$ superspaces.
These superfield equations  can also be used to search for
superfield solutions
of the superfield supergravity equations \footnote{Note 
that a toy  model of superfield interacting system,
the  coupled system of $D=2$ supergravity and a superparticle, was studied in
\cite{Knutt}. We thank
W. Kummer and A. Nurmagambetov 
for pointing out this reference.}; this might open completely new
possibilities.

On the other hand, a recent study of supergravity--superbrane interactions
indicates the gauge equivalence of the superfield description of the
dynamical supergravity interacting with a superbrane source with
a simpler system which is described by the sum of a standard
(component) supergravity action and of the action of purely bosonic
brane (a purely bosonic limit of the associated superbrane).
Until now this had been checked by a straightforward study of the $D=4$
supergravity--superparticle interacting system \cite{BAIL2}.
The present investigation shows the same result for the
supergravity---superstring system and, thus, provides  an explicit check
of such a gauge equivalence  for the dynamical system including an extended
object as a source.
This gauge equivalence also promises to be useful to search for
new solitonic solutions of superfield supergravity, with nontrivial
fermionic fields. (Very few such solutions
are known; an example is the pp--wave solution of \cite{Hull}).

This paper is organized as follows. Sec. I is devoted to the Green--Schwarz
superstring
in curved  $D=4$, $N=1$ superspace.
The superstring action (\ref{0GSac}) involves, in addition to
(the pull--backs of) the supervielbein, the scalar superfield $\Phi$ and the
two--superform $B_2$, neither of which is 
involved in the superfield description of
minimal $D=4$, $N=1$ superfield supergravity.
The flat superspace limit of the superstring action is reviewed in Sec. IA.
In Sec. IB we show (in a way close to \cite{G88})
that the requirement of  preservation of the  superstring
$\kappa$--symmetry
in the minimal curved  $D=4$, $N=1$ superspace results in  constraints for
the field strengths $H_3=dB_2$ and find that these constraints express
$H_3$ in terms of the dilaton superfield $\Phi$. Furthermore, a 
 study of the Bianchi identities $dH_3\equiv 0$ in a way close to
the one of Ref. \cite{G80,G88} (Sec. IC) concludes that
the superfield $e^{\Phi/2}$ obeys the defining constraints of the tensor
multiplet, Eqs. (\ref{0Phi=L}).

Sec. II collects the
necessary information about the superfield supergravity action
\cite{WZ78,BW} 
and their variations \cite{WZ78,BAIL2}. Sec. III describes the
admissible variations of the tensor multiplet (dilaton superfield $\Phi$)
and of the $B_2$ superform. In Sec. VI we present the complete
superfield action for the
{\sl supergravity---tensor multiplet---superstring}
interacting
system  and derive the superfield equations of motion by its
variation (Secs. IVA,B,C). [Superfield equations which follow 
from the sum of the superfield action 
of the minimal supergravity and the tensor multiplet, without 
including the superstring action, were considered in 
\cite{S95,BS96} in connection with the $D=4$ $N=1$ limit/compactification 
of the heterotic superstring].  
The superfield generalizations of the Einstein and Rarita--Schwinger
equations {\sl with sources} 
are presented in Sec. VD and those of the Kalb--Ramond
equations for tensor gauge fields in Sec. VE.

Then, in Sec. VA, by studying the gauge symmetries and using 
known results about fixing the Wess--Zumino gauge (see \cite{BAIL2} and refs.
therein) we show that the superfield description of the
{\sl supergravity---tensor multiplet---superstring}
interacting system is gauge equivalent to a
{\sl supergravity---tensor multiplet---{\it bosonic} string}
dynamical system described
by the sum of the {\it component} (spacetime) action for supergravity
interacting with tensor multiplet \cite{dWR} and the action for the
purely bosonic string, the purely bosonic limit of the Green--Schwarz
superstring. In Sec. VB we check this gauge equivalence at the level of
equations of motion by proving that the dynamical equations of motion
which follow from the complete superfield description, when considered in the
special `fermionic unitary gauge' ($\hat{Z}^M(\xi):=
(\hat{x}^\mu(\xi), \hat{\theta}^{\check{\alpha}}(\xi))=
(\hat{x}^\mu(\xi), 0)$) have the same properties as the equations
derived from the gauge fixed action (the action in
the `fermionic unitary gauge'). Namely we show that
all the dynamical equations for fermions become sourceless
in this gauge. Our conclusions are collected in Sec. VI.

\section{Green--Schwarz
superstring in a $D=4$, $N=1$ supergravity background}
\renewcommand{\theequation}{\arabic{section}.\arabic{equation}}
\setcounter{equation}0

The Green--Schwarz superstring spans a two--dimensional worldsheet
${\cal W}^2$  in superspace ${\Sigma}^{(4|4)}$,
\begin{eqnarray}\label{W2}
{\cal W}^2 \subset {\Sigma}^{(4|4)} \; , & \quad
Z^M= \hat{Z}^M(\xi^m) \; .
\end{eqnarray}
In (\ref{W2}) $Z^M= (x^\mu , \theta^{\check{\alpha}})$ are
coordinates of $D=4$, $N=1$ superspace ($\mu=0, 1,2,3$, $
\check{\alpha}=1,2,3,4$), $\xi^m= (\tau , \sigma)$ are local
worldsheet coordinates ($m=0, 1$) and $ \hat{Z}^{ {M}}(\xi )=
(\hat{x}^{ {\mu}}(\xi), \, \hat{\theta}^{\breve{\alpha}}(\xi) \,)$
are  {\it supercoordinate functions} that  define  the surface
${\cal W}^2$ in  ${\Sigma}^{(4|4)}$. One can also say that
 $\hat{Z}^{ {M}}(\xi)$ are defined by the map
 \begin{eqnarray}\label{W1s}
\hat{\phi}: W^2 \rightarrow {\Sigma}^{(4|4)}\; ,
\;\; \xi^m \mapsto \hat{Z}^{ {M}}(\xi )=
(\hat{x}^{ {\mu}}(\xi), \, \hat{\theta}^{\breve{\alpha}}(\xi)
\,) \; ,
\end{eqnarray}
of the coordinate chart  $W^2$ into ${\Sigma}^{(4|4)}$.

The $D=4$, $N=1$ version of the Green--Schwarz superstring action reads
\begin{eqnarray}\label{GSac}
& S_{sstr} & = \int_{W^2} \hat{{\cal L}}_2
= \int_{W^2} [{1 \over 4} e^{{\hat{\Phi}\over 2}}
\ast \hat{E}^a \wedge
  \hat{E}^b \eta_{ab} - \hat{B}_2] \;  \qquad
\nonumber \\ &&
\equiv  \int d^2\xi \sqrt{\det{|g|}} - \int_{W^2}\hat{B}_2\; , \qquad
\\ \label{gmn}
&& g=det(g_{mn}) \; , \qquad g_{mn} = \hat{E}^a_m \hat{E}_{na}\; .
\end{eqnarray}
It involves  the pull--backs of the forms on superspace to ${\cal
W}^2$,
\begin{eqnarray}\label{hatEa}
  \hat{E}^{ {a}} \equiv \hat{\phi}^\ast (E^a)
= d\hat{Z}^{M}(\xi) E_{M}^{~{a}}(\hat{Z})
\equiv  d\xi^m \hat{E}^{ {a}}_m \; ,
\\
\nonumber
\hat{E}^{ {a}}_m :=
\partial_m \hat{Z}^{M} E_{M}^{~ {a}}(\hat{Z}(\xi))
\;\end{eqnarray} for the bosonic supervielbein form $E^a$ on
${\Sigma}^{(4|4)}$,
\begin{eqnarray}\label{4EA}
& E^{A} \equiv ( E^a, E^{\underline{{\alpha}}})
= ( E^a, E^{{\alpha}}, \bar{E}_{{\dot\alpha}})
\quad \; ;  \\ \label{4Ea} & \qquad
E^a= dZ^M E_M^a(Z)\; , \quad
 \quad  \\ & \label{4Eal}
E^{\underline{{\alpha}}} = dZ^M E^{\underline{{\alpha}}}_M(Z)
\quad \leftrightarrow \quad \begin{cases}
E^{{\alpha}}= dZ^M E_M^{{\alpha}}(Z)\; , \\
 \bar{E}^{\dot\alpha}= dZ^M \bar{E}_M^{\dot\alpha}(Z)\; .
 \end{cases}
\end{eqnarray}
In Eqs. (\ref{4EA})--(\ref{4Eal}) $a=0, 1,2,3$ is a tangent space vector
index, ${\underline{{\alpha}}}=1,2,3,4$ is a Majorana spinor index
and ${\alpha}=1,2$, $\dot{\alpha}=1,2$ are Weyl spinor indices.

 The action (\ref{GSac}) involves also
the pull--back  $\hat{\Phi}\equiv \hat{\phi}^\ast (\Phi) =\Phi(\hat{Z})$
of a dilaton superfield $\Phi(Z)$ and the pull--back
\begin{eqnarray}
\label{hatB2}
\hat{B}_2 &\equiv \hat{\phi}^\ast (B_2) =
B_2(\hat{Z}(\xi))={1 \over 2} \hat{E}^B \wedge \hat{E}^A
B_{AB} (\hat{Z}(\xi)) \nonumber \\ & \equiv {1 \over 2}d\xi^m \wedge d\xi^n
\hat{B}_{nm}(\xi) \;  ,
\\
\nonumber &
\hat{B}_{nm}(\xi) := \partial_m \hat{Z}^{M} \partial_n \hat{Z}^{N}
B_{NM}(\hat{Z})\;
\end{eqnarray}
of a two--form on ${\Sigma}^{(4|4)}$,
\begin{eqnarray}
\label{B2}
{B}_2 ={1 \over 2} {E}^B \wedge {E}^A
B_{AB} ({Z}) \; .
\end{eqnarray}
The worldsheet Hodge star operator $\ast$
\begin{eqnarray}
\label{Hodge*}
\ast \hat{E}^a = d\xi^n \sqrt{|g|} \epsilon_{nk}g^{km}\hat{E}_m^a \; ,
\nonumber \\
\Rightarrow \quad \ast \hat{E}^a \wedge \hat{E}^b = d^2\xi
\sqrt{|g|} g^{mn} \hat{E}^a_m\hat{E}^b_n \; ,
\end{eqnarray}
can be defined using the induced worldsheet metric (\ref{gmn}).
Then
\begin{eqnarray} \label{*EE}
& {1\over 4} \ast \hat{E}_a \wedge
\hat{E}^a= {1\over 2} d^2\xi \sqrt{|g|} \; ,
\quad \\
\label{*EvE}
& \delta (\ast  \hat{E}_a \wedge \hat{E}^a) = 2 \ast \hat{E}_a \wedge \delta
\hat{E}^a \; .
\end{eqnarray}
Substituting (\ref{*EE}) in (\ref{GSac})  one arrives at the more familiar
form of the Green--Schwarz superstring action (\ref{0GSac}).

\subsection{Superstring $\kappa$--symmetry in flat superspace}

{\sl In flat superspace} one may consider a vanishing dilaton
superfield, $\Phi(Z)=0\;$ (although this is not obligatory in
$D=4$, $N=1$  superspace, where $\Phi(z)$ is described by a
separate multiplet, see below) and use  the standard expression
for the   supervielbein
\begin{eqnarray}
\label{Eaflat}
& flat\; superspace\; : \nonumber \\
& E^a = dX^\mu \delta_\mu^a -
i d\theta^\alpha \sigma^a_{\alpha\dot{\alpha}}
\bar{\theta}^{\dot{\alpha}} +
i \theta^\alpha \sigma^a_{\alpha\dot{\alpha}}
d\bar{\theta}^{\dot{\alpha}} 
\; ,  \\
\label{Ealflat}
& E^\alpha = d\theta^{\check{\beta}}\delta_{\check{\beta}}{}^\alpha \equiv
d\theta^\alpha\; , \nonumber \\
& \bar{E}^{\dot\alpha} = d\theta^{\check{\beta}}\delta_{\check{\beta}}
{}^{\dot\alpha}
 \equiv d\bar{\theta}^{\dot\alpha}\; ,
\end{eqnarray}
where $\check{\alpha}=1,2,3,4$ shall be treated as a Majorana spinor index,
$\check{\alpha}=\underline{\alpha}\, $;
$\theta^{\check{\beta}}=
\theta^{\underline{\beta}} = (\theta^{{\beta}}, \bar{\theta}_{\dot{\beta}})$
({\it i.e.} $\theta^{\check{\beta}}\equiv
\theta^{{\alpha}}\delta^{\check{\beta}}
{}_{\alpha} + \bar{\theta}_{\dot{\alpha}}\delta^{\check{\beta}\dot\alpha}$).
The `vacuum value' of the two--form $B_2$ has to be chosen as
\begin{eqnarray}
\label{B20}
& flat\; superspace\; , \;\; \Phi(Z)=0\; : \nonumber \\
& {B}_2 =
- {i\over 2} dX^\mu \delta_\mu^a\wedge
[d\theta^\alpha \sigma_{a\alpha\dot{\alpha}}
\bar{\theta}^{\dot{\alpha}}-
\theta^\alpha \sigma_{a\alpha\dot{\alpha}}
d\bar{\theta}^{\dot{\alpha}}]\; . \quad 
\end{eqnarray}
Then the pull--back $\hat{B}_2={\phi}^\ast (B_2)$ of the two--form
$B_2$ (\ref{B20}) produces the standard form of the Wess--Zumino
term of the $D=4$ Green--Schwarz superstring action \cite{GS84}.
Due to its presence the action possesses a local fermionic
symmetry, the seminal  $\kappa$--symmetry \cite{ALS,GS84}. Its
transformation rules can be formulated as follows
\begin{eqnarray}
\label{kappaG}
i_\kappa \hat{E}^a := \delta_{\kappa} \hat{Z}^M E_M^a(\hat{Z})=0 \; ,
\qquad \nonumber \\
i_\kappa \hat{E}^\alpha \sigma_{a\alpha \dot{\alpha}} \,
(\ast \hat{E}^a- \hat{E}^a) = 0 \, . \qquad
\end{eqnarray}
Indeed, as $(\ast \hat{E}^a- \hat{E}^a)= d\xi^n (\delta_n^m + \sqrt{|g|}
\epsilon_{nk}g^{km}) \hat{E}^a_m$ and, in flat superspace,
$E^a$ has the form of Eq. (\ref{Eaflat}),
the solution of Eqs. (\ref{kappaG})
with respect to $\delta_{\kappa}X^\mu$ and $\delta_{\kappa}\theta^\alpha$
gives the $D=4$ version of the
Green--Schwarz expression for the superstring
$\kappa$--symmetry \cite{GS84},
\begin{eqnarray}
\label{kappa0}
& flat\; superspace\;  , \;\; \Phi(Z)=0\; : \quad \nonumber \\
& \delta_{\kappa}\hat{X}^\mu=
i \delta_{\kappa}\theta^\alpha \sigma^a_{\alpha\dot\alpha}
\bar{\theta}^{\dot\alpha} + c.c. \; , \nonumber
\\
& \delta_{\kappa} \hat{\theta}^\alpha = \bar{\kappa}^n_{\dot\alpha}
(\delta_n^m - \sqrt{|g|}
\epsilon_{nk}g^{km}) \hat{E}^a_m \tilde{\sigma}_a^{\dot{\alpha}\alpha} \; ,
\\ & \qquad  \delta_{\kappa} \hat{\bar{\theta}}{}^{\dot\alpha}= (
\delta_{\kappa} \hat{\theta}^\alpha)^* \; . \nonumber
\end{eqnarray}

\subsection{Superstring $\kappa$--symmetry and superspace constraints
for supergravity and the tensor multiplet.}

When the Green--Schwarz superstring is considered in curved
superspace, {\it i.e.} in the presence of superfield supergravity,
the natural selfconsistency condition is the existence of a smooth
flat superspace limit. This implies, in particular, that the
number of local symmetries of the action in {\sl a superfield
supergravity background} should be the same as in the case of flat
superspace. This means, again in  particular, that 
$\kappa$--symmetry should be present in the superstring model in a
supergravity background.

However, it is well known that  $\kappa$--symmetry occurs only when
 the background satisfies certain constraints \cite{strKsg,BST87}.
This occurs also with the $D=4$ $N=1$ Green--Schwarz superstring 
\cite{G88} (see also \cite{S88,S95,G96}). 
For the sake of completeness and to establish the notation 
used in next sections 
we present here some details of the constraint derivation.

The variation of the Green--Schwarz superstring action (\ref{GSac}) in
a general curved superspace looks like
 \begin{eqnarray}\label{dGSac}
\delta S_{sstr} &=&
\int_{W^2} \left[ {1 \over 2} e^{{\hat{\Phi}\over 2}}
\ast \hat{E}_a \wedge
  \delta \hat{E}^a +  \right. \nonumber \\
&+& \left.  {1 \over 8} e^{{\hat{\Phi}\over 2}}
\ast \hat{E}_a \wedge
  \hat{E}^a \delta \Phi - \delta\hat{B}_2 \right] \; . \qquad
\end{eqnarray}
As the only dynamical variables in a curved superspace {\sl background}
are the supercoordinate functions $\hat{Z}^M(\xi)$, one only 
considers in this case
\begin{eqnarray}
\label{dZGSac}
 \delta_{\hat{Z}} S_{sstr} & = &
 \int_{W^2} \left[ {1 \over 2} e^{{\hat{\Phi}\over 2}}
\ast \hat{E}_a \wedge
  \delta_{\hat{Z}} \hat{E}^a + \right. \nonumber \\
& +& \left. {1 \over 8} e^{{\hat{\Phi}\over 2}}
\ast \hat{E}_a \wedge
  \hat{E}^a \delta_{\hat{Z}} \Phi
- \delta_{\hat{Z}}\hat{B}_2\right] \;  , \qquad
\end{eqnarray}
and the variations $\delta_{\hat{Z}}$ of the pull--backs of differential forms
are given by Lie derivatives,
 \begin{eqnarray}
\label{*gcEa}
\delta_{\hat{Z}} \hat{E}^{a} & \equiv
\delta_{\hat{Z}} {E}^{a}(\hat{Z}) :=  E^{a} (\hat{Z}+ \delta\hat{Z}) -
E^{a}(\hat{Z})=
\nonumber \\
& =  i_{\delta \hat{Z}}(d \hat{E}^{a})  + d(i_{\delta \hat{Z}}\hat{E}^{a}) =
\nonumber \\
& =  i_{\delta \hat{Z}} \hat{T}^{a} +
{\cal D}(i_{\delta \hat{Z}} \hat{E}^a) +
\hat{E}^{b} i_{\delta \hat{Z}} w_{b}{}^{a} \; ,
\\ \label{*gcB2}
&
\delta_{\hat{Z}} \hat{B}_2 = i_{\delta \hat{Z}} \hat{H}_3 +
di_{\delta \hat{Z}} \hat{B}_2\; , \qquad H_3:= dB_2 \; ,
\end{eqnarray}
where
\begin{eqnarray}
 \label{iZEa}
& i_{\delta \hat{Z}} E^{a}(\hat{Z}):= \delta \hat{Z}^M E_M^{a}(\hat{Z}) \; ,
\qquad  \\ \label{iZwab}
& i_{\delta \hat{Z}} w^{ab}:=  \delta \hat{Z}^M\; w_M^{ab}(\hat{Z})\; ,
\qquad  \\ \label{iZTa}
& i_{\delta \hat{Z}} \hat{T}^{a} := \hat{E}^C \, i_{\delta \hat{Z}}\hat{E}^B
T_{BC}{}^a(\hat{Z}) \; , \quad etc. ,
\end{eqnarray}
the torsion $T^a$ and the covariant exterior derivative ${\cal D}$
are defined below (Eqs. (\ref{4WTa=def})--(\ref{4WTdA=def})) and
$w^{ab}=dZ^M w_M{}^{ab}= - w^{ba}$ is the spin connection.

Now it is clear that $\kappa$--symmetry is not present in a
general  curved superspace. As we are interested in the
superstring interaction with supergravity, we have to impose first
the constraints on the torsion of curved superspace
\begin{eqnarray}\label{4WTa=def}
T^a & :=  {\cal D}E^a =
dE^{a} - E^b \wedge w_b {}^a\; \equiv  \qquad \nonumber \\ & \qquad
\equiv
{1\over 2} E^B \wedge E^C T_{CB}{}^a \; ,    \\
\label{4WTal=def}
T^{\alpha} & :=  {\cal D}E^{\alpha} =
dE^{\alpha} - E^\beta \wedge w_\beta {}^\alpha \equiv
\qquad \nonumber \\ & \qquad \equiv
{1\over 2} E^B \wedge E^C T_{CB}^{\alpha} \; ,
 \\
\label{4WTdA=def}
T^{\dot{\alpha}} & :=  {\cal D}\bar{E}^{\dot{\alpha}}
=
d \bar{E}^{\dot{\alpha}} - \bar{E}^{\dot{\beta}} \wedge
w_{\dot{\beta}}{}^{\dot{\alpha}} \equiv
\qquad \nonumber \\ & \qquad \equiv
{1\over 2} E^B \wedge E^C T_{CB}^{\dot{\alpha}} \; ,  
\end{eqnarray}
and on the  Riemann curvature two--form
\begin{eqnarray}\label{4WR=def0}
R^{ab}& := dw^{ab} -w^{ac}\wedge w_c{}^b
\equiv {1 \over 2} E^C \wedge E^D R_{DC}{}^{ab}
\; .
\end{eqnarray}

The {\it minimal} off--shell formulation of $D=4$, $N=1$ supergravity
is described by the set of constraints (see \cite{BW} and refs. therein)
\begin{eqnarray}\label{4DconTa}
& T_{\alpha \dot{\beta}}{}^a= -2i\sigma^a_{\alpha \dot{\beta}}\; , \nonumber
\\ \label{4DconT}
& T_{\alpha\beta}{}^A=0=T_{\dot{\alpha}\dot{\beta}}{}^A\; ,
\quad T_{\alpha\dot{\beta}}{}^{\dot{\gamma}}=0\; ,  \quad T_{\alpha b}{}^c=0
\; , \\  \label{4DconR}
& R_{\alpha\dot{\beta}}{}^{ab}=0 \; .
 \end{eqnarray}

These constraints and their consequences (derived from the Bianchi
identities) can be collected in the following expressions for the
torsion two--forms \cite{WZ78,BW}
\begin{eqnarray}\label{4WTa=}
T^a & =- 2i\sigma^a_{\alpha\dot{\alpha}} E^\alpha \wedge
\bar{E}^{\dot{\alpha}} +{1\over 16} E^b \wedge E^c
\varepsilon^a{}_{bcd} G^d \; ,
\\
\label{4WTal=}
T^{\alpha} &
= {i\over 8} E^c \wedge E^{\beta} (\sigma_c\tilde{\sigma}_d)_{\beta}
{}^{\alpha} G^d  - \hspace{2.3cm}
\nonumber \\ &  -{i\over 8} E^c \wedge \bar{E}^{\dot{\beta}}
\epsilon^{\alpha\beta}\sigma_{c\beta\dot{\beta}}R
+
 {1\over 2} E^c \wedge E^b \; T_{bc}{}^{\alpha}\; , \\
\label{4WTdA=}
T^{\dot{\alpha}}
& = {i\over 8} E^c \wedge E^{\beta} \epsilon^{\dot{\alpha}\dot{\beta}}
\sigma_{c\beta\dot{\beta}}
\bar{R} - \hspace{2.3cm}
\;
\nonumber \\ &
-{i\over 8} E^c \wedge \bar{E}^{\dot{\beta}}
(\tilde{\sigma}_d\sigma_c)^{\dot{\alpha}}{}_{\dot{\beta}}
\, G^d +
 {1\over 2} E^c \wedge E^b \; T_{bc}{}^{\dot{\alpha}}\; , \quad 
\end{eqnarray}
and for the superspace Riemann curvature 2--form
\begin{eqnarray}\label{4WR=def}
R^{ab}& := dw^{ab} -w^{ac}\wedge w_c{}^b = \qquad {} \qquad
{} \quad \nonumber \\ & = {1\over 2} R^{\alpha\beta}
(\sigma^a\tilde{\sigma}^b)_{\alpha\beta}
- {1\over 2} R^{\dot{\alpha}\dot{\beta}}
(\tilde{\sigma}^a\sigma^b)_{\dot{\alpha}\dot{\beta}} \; , 
\end{eqnarray}
\begin{eqnarray}\label{4WR=}
R^{\alpha\beta} & \equiv  dw^{\alpha\beta} - w^{\alpha\gamma}
\wedge w_\gamma{}^\beta \equiv {1 \over 4}
R^{ab} (\sigma_a\tilde{\sigma}_b)^{\alpha\beta}= \nonumber
\\
&
= -{1\over 2} E^\alpha \wedge E^\beta \bar{R}
-{i\over 8} E^c \wedge E^{(\alpha}\,
\tilde{\sigma}_c{}^{\dot{\gamma}\beta)} \bar{{\cal D}}_{\dot{\gamma}}\bar{R} -
\nonumber \\
&
+{i\over 8} E^c \wedge E^{\gamma}
(\sigma_c\tilde{\sigma}_d)_{\gamma}{}^{(\beta} {\cal D}^{\alpha)} G^d -
\nonumber \\ &
-{i\over 8}  E^c \wedge \bar{E}^{\dot{\beta}} \sigma_{c\gamma\dot{\beta}}
W^{\alpha\beta\gamma} + {1\over 2} E^d \wedge E^c R_{cd}{}^{\alpha\beta} \; ,
\quad  \\ \label{4WRcc}
& R^{\dot{\alpha}\dot{\beta}}=
(R^{{\alpha}{\beta}})^* \: .
\end{eqnarray}
The {\it r.h.s}'s of Eqs. (\ref{4WTa=})--(\ref{4WR=}) include the so--called
`main superfields' $R$, $\bar{R}=R^*$, $G_a=(G_a)^*$ and
the symmetric spin--tensor
$W_{\alpha\beta\gamma} =W_{(\alpha\beta\gamma)}=
(\bar{W}_{\dot{\alpha}\dot{\beta}\dot{\gamma}})^*$,
which obey (also as a result of Bianchi identities)
\begin{eqnarray}
\label{chR}
& {\cal D}_\alpha \bar{R}=0\; , \qquad \bar{{\cal D}}_{\dot{\alpha}} {R}=0\; ,
 \\
\label{chW} & \bar{{\cal D}}_{\dot{\alpha}} W^{\alpha\beta\gamma}= 0\; ,
\qquad
{{\cal D}}_{{\alpha}} \bar{W}^{\dot{\alpha}\dot{\beta}\dot{\gamma}}= 0\; ,
\\
\label{DG=DR} &
\bar{{\cal D}}^{\dot{\alpha}}G_{{\alpha}\dot{\alpha}}= {\cal D}_{\alpha} R
\; , \qquad
{{\cal D}}^{{\alpha}}G_{{\alpha}\dot{\alpha}}=
\bar{{\cal D}}_{\dot{\alpha}} R \; ,
\\ \label{DW=DG} &
{{\cal D}}_{{\gamma}}W^{{\alpha}{\beta}{\gamma}}=
\bar{{\cal D}}_{\dot{\gamma}} {{\cal D}}^{({\alpha}}G^{{\beta})\dot{\gamma}}
\; , \qquad \nonumber \\ &
\bar{{\cal D}}_{\dot{\gamma}} \bar{W}^{\dot{\alpha}\dot{\beta}\dot{\gamma}}
= {{\cal D}}_{{\gamma}} \bar{{\cal D}}^{(\dot{\alpha}|}
G^{{\gamma}|\dot{\beta})} \; . \qquad
\end{eqnarray}
Here and below the brackets (square brackets) denote
symmetrization (antisymmetrization) with  unit weight, {\it e.g.}
${{\cal D}}^{({\alpha}}G^{{\beta})\dot{\gamma}}:= {1\over 2}
({{\cal D}}^{\alpha}G^{\beta\dot{\gamma}} + {{\cal
D}}^{\beta}G^{\alpha\dot{\gamma}})$.

After the minimal $D=4$, $N=1$ supergravity constraints
(\ref{4WTa=}), (\ref{4WTal=}) are taken into account, one finds for the
variation of the superstring action
 \begin{eqnarray}\label{dZGSac4}
& \delta_{\hat{Z}}S_{sstr}  = \hspace{6cm} \nonumber \\
&  - {1 \over 2} \int_{W^2}
[{\cal D}(e^{{\hat{\Phi}\over 2}}* \hat{E}_a ) -
{1 \over 4} e^{{\hat{\Phi}\over 2}}
* \hat{E}_b \wedge \hat{E}^b \hat{\nabla}_a \hat{\Phi}]
i_{\delta \hat{Z}}E^a - \,   \nonumber  \\
& - i \int_{W^2} e^{{\hat{\Phi}\over 2}} * \hat{E}_a\wedge
\sigma_{a\alpha\dot{\alpha}} \hat{\bar{E}}{}^{\dot\alpha} + \, c.c.\, -
\nonumber \\
& \quad -
{i\over 8} \int_{W^2}
e^{{\hat{\Phi}\over 2}}
* \hat{E}_b \wedge \hat{E}^b \hat{\nabla}_{\alpha}\hat{\Phi}\,
i_{\delta \hat{Z}}E^\alpha + c.c. - \; \nonumber \\
& \hspace{2cm} -  \int_{W^2}i_{\delta \hat{Z}}H_3 \; ,
\end{eqnarray}
where we denote $\nabla_A {\Phi}\vert_{Z=\hat{Z}(\xi)}
:= \hat{\nabla}_A \hat{\Phi}$ and
 ignore the boundary contribution
$\int_{W^1= \partial W^2} [{1\over 2}e^{{\hat{\Phi}\over 2}}
* \hat{E}_b \, i_{\delta \hat{Z}}\hat{E}^b - i_{\delta \hat{Z}}\hat{B}_2]$
(which  always vanishes for the case of a closed superstring with
$\partial W^2 = \emptyset$). The above consideration of the flat
superspace case suggests that  $\kappa$--symmetry occurs in the
action (\ref{GSac}) when also the constraints
 \begin{eqnarray} \label{Hc00}
H_{\alpha \beta \gamma}=0\; , \qquad H_{\alpha \beta \dot{\gamma}}=0\; ,
\qquad and \quad c.c. \nonumber \\
H_{\alpha \beta c}=0\; , \qquad
H_{\dot{\alpha} \dot{\beta} c}=0\; ,
\\ \label{Hc0}
H_{\alpha \dot{\alpha} a}= - i e^{{\hat{\Phi}\over 2}}
\sigma_{a \alpha \dot\alpha}
\qquad
\end{eqnarray}
are imposed on the field strength $H_3=dB_2$ of the two--superform
$B_2$ (\ref{B2}).

\subsection{Gauge superform and tensor multiplet}

The  study of the Bianchi identities $dH_3\equiv 0$ in the
superspace restricted by the supergravity constraints
(\ref{4WTa=}), (\ref{4WTal=}), (\ref{4WTdA=}) shows that the field
strength $H_3$ is completely determined by the constraints
(\ref{Hc00}), (\ref{Hc0}). It is expressed in terms of the dilaton
superfield $\Phi (Z)$ and the main superfields of the minimal
supergravity by
\begin{eqnarray}\label{H3}
H_3 &\equiv& dB_2  = - i E^a \wedge E^{\alpha} \wedge E^{\dot\alpha}
\sigma_{a \alpha  \dot\alpha}
e^{{{\Phi}\over 2}}  + \qquad \nonumber \\
& +& {1\over 8} E^a \wedge E^b \wedge  E^{\alpha}
(\sigma_{[a}\tilde{\sigma}_{b]})_{\alpha}{}^{\beta} e^{{\Phi\over 2}}
\nabla_{\beta} \Phi  + c.c.
+ \nonumber \\
& +& {1\over 3!}  E^a \wedge E^b \wedge  E^c H_{cba} \: ,
\\ \label{Habc}
 H_{abc} & =& {5\over 32} e^{{\Phi\over 2}} \epsilon_{abcd}G^d
+  {1\over 8} \epsilon_{abcd} \tilde{\sigma}^{d\dot\alpha\alpha}
[{\cal D}_{\alpha}, \bar{{\cal D}}_{\dot{\alpha}}] e^{{\Phi\over 2}} \; . 
\qquad 
\end{eqnarray}

Thus the two form field strength is expressed, essentially, through
one real dilaton superfield $\Phi(Z)$.

Moreover, the same study of Bianchi identities brings also the
equations which, on first sight, seem to be relations between
the dilaton superfield and the chiral main superfield of minimal
supergravity
\begin{eqnarray}\label{RPhi}
\bar{R}= e^{-{\Phi\over 2}}{\cal D}\nabla e^{{\Phi\over 2}} \; , \qquad
{R}= e^{-{\Phi\over 2}}\bar{{\cal D}}\bar{\nabla} e^{{\Phi\over 2}} \; .
 \end{eqnarray}
However, one notes that Eqs. (\ref{RPhi}) can be written as
\begin{eqnarray}\label{Phi=lnL}
({\cal D}{\cal D} -
\bar{R}) e^{{\Phi\over 2}}=0 \; , \qquad
(\bar{{\cal D}}\bar{{\cal D}} - R)e^{{\Phi\over 2}} = 0 \; ,
 \end{eqnarray}
and they just imply that the $D=4$, $N=1$ dilaton superfield describes
the real linear multiplet\footnote{One should keep in mind that
$({\cal D}{\cal D} - \bar{R})$ is a chiral projector, {\it i.e.}
that ${\cal D}_\alpha\, ({\cal D}{\cal D} - \bar{R}) U \equiv 0$
for any superfield $U=U(Z)$.}.

The fact that a two--form in $D=4$, $N=1$ superspace is described
by real linear multiplet (tensor multiplet) is known from the
study in \cite{G80}.

Thus we conclude that the complete superfield action for the
$D=4$, $N=1$ interacting system including the superstring should
involve, in addition to the superstring action (\ref{GSac}) and
the Wess--Zumino action for the {\sl minimal} supergravity
multiplet, $\int d^8 Z \, E\, $ (see Eq. (\ref{SGact}) below) also
an action for the tensor multiplet, described by the real superfield
$e^{{\Phi\over 2}}$ which obeys the constraints (\ref{Phi=lnL}).
The kinetic term of the latter action should involve, in
particular, the kinetic term for the two--index antisymmetric
tensor gauge field (or Kalb--Ramond field  
\cite{KR}, first introduced in $D=4$ 
under the {\it notoph} name \cite{OP}.) which interacts naturally
with a string. Such a kinetic term can be written as \cite{dWR}
\begin{eqnarray}\label{ITMS}
\int d^8 Z \, E \, {\Phi\over 2} e^{{\Phi\over 2}}\; .
\end{eqnarray}
Note that the first proposal for the tensor multiplet action was
different, $\int d^8 Z \, E \, e^{\Phi} \equiv \int d^8 Z \, E \,
(e^{{\Phi\over 2}})^2$ (see \cite{G80} and refs. therein). The
tensor multiplet with the action (\ref{ITMS}) was referred to as
'improved tensor multiplet' \cite{dWR}. Its distinguishing
property is invariance under the Weyl transformations acting
also on the dilaton superfield, $E^a\rightarrow e^{\Lambda} E^a$,
$E^{\underline{\alpha}}\rightarrow e^{\Lambda/2
}E^{\underline{\alpha}}$, $\Phi(z) \rightarrow \Phi (Z) - 4
\Lambda$, when $\Lambda$ is given by a sum of chiral and
antichiral superfields, $\Lambda=i(\phi -\bar{\phi})$, ${\cal
D}_\alpha \bar{\phi}= 0 = \bar{{\cal D}}_{\dot{\alpha}}{\phi}$.
The fact that the superstring action (\ref{GSac}) also possesses
such a symmetry makes the improved tensor multiplet action
preferable for a description of the tensor multiplet--superstring
interacting system.

Actually in the study of $D=4$ $N=1$ limit/compactification  of 
the heterotic string \cite{4Dlimit,S88,G88,S95,G96,BS96} it was argued
that such a limit rather provided the  
{\sl minimal supergravity---tensor multiplet} action \cite{S88,G88}. 
We are not addressing this problem here but rather considering the 
$D=4$ $N=1$ interacting system 
(including the $D=4$ $N=1$ superstring) as a 
relatively simple 
model for the (quasi)classical description of a more complicated, higher 
dimensional ($D=10, 11$) supergravity---superbrane interacting system 
[in particular, the $D=10$ $N=1$ 
{\sl supergravity--super Yang--Mills---heterotic string} interacting system 
described by a hypothetical superfield action also including, in addition to 
the supergravity and the super--YM parts, the heterotic string action 
\cite{Heterotic}, as well as the 
$D=10$ {\sl type II supergravity---super--D$p$--brane} 
and $D=11$ {\sl supergravity---super--Mp--brane} systems]. As so, 
an interesting alternative possibility is to consider the
so--called {\sl new minimal} formulation of the simple
supergravity \cite{SW81}, where the {\sl auxiliary} fields can be
collected into a real linear multiplet. Here we will not consider
this possibility, but proceed with the description of superstring
interacting with dynamical real linear multiplet and minimal $N=1$
supergravity \footnote{In the framework of the conformal tensor
calculus, to arrive at the {\sl component} form of the action of
the new minimal (off-shell) supergravity one starts from the
improved tensor multiplet action in flat superspace, $\int d^8Z \,
L\, ln L$ with $DDL=0=\bar{D}\bar{D}L$, performs the Grassmann
integration in it, then one introduces the coupling to the conformal
supergravity and makes a gauge choice for the special
superconformal transformation. As a result, one arrives at the
action  \cite{SW81} $\int d^4 x (e {\cal R} +
\epsilon^{\mu\nu\rho\sigma} \bar{\psi}_\mu\gamma_5\gamma_\nu {\cal
D}_\rho \psi_\sigma + A_\mu \epsilon^{\mu\nu\rho\sigma}
\partial_\nu B_{\rho\sigma} + {1\over 2}
(\epsilon^{\mu\nu\rho\sigma} \partial_\nu B_{\rho\sigma})^2)$,
which includes both the antisymmetric tensor $B_{\mu\nu}$ and the
vector gauge field $A_\mu$ as {\it auxiliary fields} (see
\cite{dWR}). In contrast, in our case the interacting action
(\ref{0Sint}) involves the improved tensor multiplet action
coupled to minimal supergravity in superfield formulation, $\int
d^8Z \, E\, L\, ln L$ with $ln L= \Phi/2$. In this case the
antisymmetric tensor field $B_{\mu\nu}$ is dynamical and its
equations of motion contain a source from the superstring.}.

\subsection{Superstring $\kappa$--symmetry in supergravity and tensor
multiplet background}

With the constraints (\ref{4WTa=}), (\ref{4WTal=}), (\ref{4WTdA=}) and
(\ref{H3}) the variation (\ref{dZGSac4}) of the superstring action (\ref{GSac})
becomes
 \begin{eqnarray}\label{dZGSacf}
& \delta_{\hat{Z}}S_{sstr}  = \hspace{6cm} \nonumber \\
&  - {1 \over 2} \int_{W^2}
[{\cal D}(e^{{\hat{\Phi}\over 2}}* \hat{E}_a ) -
{1 \over 4} e^{{\hat{\Phi}\over 2}}
* \hat{E}_b \wedge \hat{E}^b \hat{\nabla}_a \hat{\Phi}]
i_{\delta \hat{Z}}E^a - \,   \nonumber  \\
& + i \int_{W^2} \, e^{{\hat{\Phi}\over 2}} (\hat{E}_a- \ast \hat{E}_a)\wedge
\sigma_{a\alpha\dot{\alpha}} \hat{\bar{E}}{}^{\dot\alpha} + c.c. +\nonumber \\
& \quad +
{i\over 8} \int_{W^2} 
e^{{\hat{\Phi}\over 2}}
(\hat{E}_b- \ast\hat{E}_b) \wedge \hat{E}^b \hat{\nabla}_{\alpha}\hat{\Phi}\, 
i_{\delta \hat{Z}}E^\alpha + c.c. \qquad 
\end{eqnarray}
Eq. (\ref{dZGSacf}) makes evident the presence of the local fermionic
$\kappa$--symmetry defined by Eq. (\ref{kappaG}), but now with
curved space supervielbein,
\begin{eqnarray}
\label{kappaSG}
i_\kappa \hat{E}^a =0 \; ,
\qquad
i_\kappa \hat{E}^\alpha \sigma_{a\alpha \dot{\alpha}} \,
(* \hat{E}^a- \hat{E}^a) = 0 \, .  \qquad
\end{eqnarray}
The solution of Eqs. (\ref{kappaSG})
({\it cf.} Eq. (\ref{kappa0}) and above) provides us with the explicit
form of the $\kappa$--symmetry transformations,
\begin{eqnarray}
\label{kappaSGZ}
\delta_{\kappa} \hat{Z}^M(\xi) &=& \bar{\kappa}^n_{\dot\alpha}
(\delta_n^m - \sqrt{|g|}
\epsilon_{nk}g^{km}) \hat{E}^a_m \tilde{\sigma}_a^{\dot{\alpha}\alpha}
\, E_{\alpha}^M(\hat{Z}) +  \nonumber \\ &&  + \, c.c.   \; ,
\end{eqnarray}
where $g^{nm}(\xi)$ is the matrix inverse to the induced metric
\begin{eqnarray}
\label{gind}
g_{mn}(\xi)&=& E_m^a(\hat{Z})   E_m^b(\hat{Z})\eta_{ab}=
\nonumber \\ & = &
\partial_m  \hat{Z}^M(\xi)
 \partial_n  \hat{Z}^N(\xi)\, E_{Na}(\hat{Z})\, E_M^a(\hat{Z})\; . \qquad
\end{eqnarray}
The standard flat superspace Green--Schwarz $\kappa$--symmetry
transformations (\ref{kappa0}) can be derived from
(\ref{kappaSGZ}) by substitution of the flat superspace
expressions for the (inverse) supervielbein coefficients
$E_{\alpha}^M(\hat{Z})$ and for $ E_M^a(\hat{Z})$ in (\ref{gind}).

\section{$D=4$ $N=1$ superfield supergravity action}
\renewcommand{\theequation}{\arabic{section}.\arabic{equation}}
\setcounter{equation}0

The action of $D=4$ $N=1$ supergravity is given by the invariant supervolume
of $D=4$, $N=1$ superspace \cite{WZ78}
\begin{eqnarray}\label{SGact}
S_{SG} = \int d^4 x \tilde{d}^4\theta \; sdet(E_M^A) \;
\equiv \int d^8Z \; E \; ,
\end{eqnarray}
where $E:= sdet(E_M^A)$ is Berezinian (superdeterminant) of the
supervielbein $E_M^A(Z)$, Eq. (\ref{4Ea}),  and
$E_M^A(Z)$ are assumed to be subject to the constraints
(\ref{4DconTa}), (\ref{4DconR}).

\subsection{Admissible variations of supervielbein}

As the supervielbein is considered to be restricted by the
constraints, its variation cannot be treated as
independent\footnote{Note that with an independent variation of
the supervielbein, the action (\ref{SGact}) would lead to the equation
stating the vanishing of the superdeterminant $E$, which
contradicts  the original assumption about its nondegeneracy.}.
To find admissible variation one can, following \cite{WZ78},
denote the general variation of the supervielbein and spin
connections by
\begin{eqnarray}\label{varEMA}
\delta E_M^{\, A}(Z) = E_M^{\, B} {\cal K}_{B}^{\, A}
(\delta )\; , \quad
\delta w_M^{ab}(Z) = E_M^{\, C} u_{{C}}^{ab} (\delta )\; ,
\end{eqnarray}
and obtain the  equations to be satisfied by
${\cal K}_{B}^{\, A}  (\delta )$,
$u_{C}^{ab} (\delta )$ from the requirement
that the constraints (\ref{4DconTa}), (\ref{4DconR}) are preserved
under (\ref{varEMA}).

Quite complicated but straightforward calculations result in the
following expression \cite{BAIL2} for admissible variations of
the supervielbein
\begin{eqnarray}\label{varEa}
\delta E^{a} & = E^a (\Lambda (\delta ) + \bar{\Lambda} (\delta ))
 - {1\over 4} E^b \tilde{\sigma}_b^{ \dot{\alpha} {\alpha} }
[{\cal D}_{{\alpha}}, \bar{{\cal D}}_{\dot{\alpha}}] \delta H^a +
 \hspace{-0.5cm}
\nonumber \\
& + i E^{\alpha} {\cal D}_{{\alpha}}\delta H^a
- i \bar{E}^{\dot{\alpha}}\bar{{\cal D}}_{\dot{\alpha}} \delta H^a \; ,
\\ \label{varEal}
 \delta E^{\alpha} & =  E^a \Xi_a^{\alpha}(\delta ) +
E^{\alpha} \Lambda (\delta )
+ {1\over 8} \bar{E}^{\dot{\alpha}} R \sigma_a{}_{\dot{\alpha}}{}^{\alpha}
\delta H^a \; .
\end{eqnarray}
In Eqs. (\ref{varEa}), (\ref{varEal}),
$\Lambda (\delta )$, $\bar{\Lambda} (\delta )$ are given by
\begin{eqnarray}\label{Lb}
\Lambda (\delta ) & = {1\over 24}
\tilde{\sigma}_a^{ \dot{\alpha} {\alpha} }
[{\cal D}_{{\alpha}}, \bar{{\cal D}}_{\dot{\alpha}}] \delta H^a +
{i\over 4}{\cal D}_a  \delta H^a + {1\over 24} G_a   \delta H^a
\nonumber \\ & + 2 ( {\cal D}{\cal D}- \bar{R})\delta {\cal U}
- ( \bar{{\cal D}}\bar{{\cal D}}- {R})\delta \bar{{\cal U}} \;
\\ \label{Lb+cc}
\Lambda (\delta ) & + \bar{\Lambda} (\delta )  =
{1\over 12}
\tilde{\sigma}_a^{ \dot{\alpha} {\alpha} }
[{\cal D}_{{\alpha}}, \bar{{\cal D}}_{\dot{\alpha}}] \delta H^a +
{1\over 12} G_a   \delta H^a + \nonumber \\
& + ( {\cal D}{\cal D}- \bar{R})\delta {\cal U}
+ ( \bar{{\cal D}}\bar{{\cal D}}- {R})\delta \bar{{\cal U}} \; ;
\end{eqnarray}
the explicit expression for $\Xi_a^{\alpha}(\delta )$
in (\ref{varEal}) as well as the expression for the basic variations
of the spin connection, $u_{{C}}^{ab} (\delta )$ in Eq. (\ref{varEMA}),
will not be needed below
(they can be found in \cite{BAIL2}).

Note that the variations representing the manifest gauge symmetries
of supergravity are factored out from the above expressions. These are
the superspace local Lorentz transformations and the variational version of
the superspace general coordinate transformations (see \cite{WZ78}).

For the free supergravity action  (\ref{SGact})
the nontrivial dynamical equations of motion should follow from the
variations (\ref{varEa}), (\ref{varEal}) with (\ref{Lb}), (\ref{Lb+cc})
only.
The variation of the superdeterminant $E=sdet(E_M^A)$
under  (\ref{varEa}), (\ref{varEal}), has the form
(see \cite{WZ78})
\begin{eqnarray}\label{varsdE}
\delta E = & E [ - {1\over 12} \tilde{\sigma}_a^{\dot{\alpha}\alpha} [
{\cal D}_{\alpha}, \bar{\cal D}_{\dot{\alpha}}] \delta H^a +
{1\over 6} G_a \; \delta H^a +
 \nonumber \\
& +  {2}
(\bar{\cal D} \bar{\cal D} - R) \delta \bar{{\cal U}}+  {2}
({\cal D} {\cal D} - \bar{R}) \delta {\cal U}] \; .
\end{eqnarray}
In the light of the identity
\begin{eqnarray}
\label{idd8}
& \int d^8Z E \; {\cal D}_{A}\xi^{A}(-1)^{A}\;
= \qquad \nonumber \\ & =   \int d^8Z
E \; ({\cal D}_{A}\xi^{A} + \xi^B T_{BA}{}^A) (-1)^{A}\; \equiv 0 \; ,
 \end{eqnarray}
all the terms with derivatives can be omitted in (\ref{varsdE})
 when one considers the variation of the action (\ref{SGact}).
[The first equation in  (\ref{idd8}) uses the minimal
supergravity constraints which imply $T_{BA}{}^A) (-1)^A=0$].
Hence,
\begin{eqnarray}\label{vSGsf}
& \delta S_{SG} = \int d^8 Z \; \delta E = \nonumber \\ & =
\int d^8Z E\;  [{1\over 6} G_a \; \delta H^a -
{2}  R\; \delta \bar{{\cal U}}
-  {2}   \bar{R}\; \delta {\cal U}] \;
\end{eqnarray}
and one arrives at the following {\sl superfield equations of
motion for `free', simple $D=4$, $N=1$ supergravity}:
\begin{eqnarray}\label{SGeqmG}
 {\delta S_{SG}\over \delta H^a}=0 \quad \Rightarrow \quad  G_a =0 \; ,
\\ \label{SGeqmR}
{\delta S_{SG}\over \delta \bar{{\cal U}}}=0 \quad \Rightarrow \quad
 R=0 \; ,
\\ \label{SGeqmbR}
{\delta S_{SG}\over \delta {{\cal U}}}=0 \quad \Rightarrow
\quad \bar{R}=0 \; .
\end{eqnarray}

\section{Tensor multiplet in curved superspace}
\setcounter{equation}0
\subsection{The `improved' action for tensor multiplet }

As was argued in Sec. ID (see also  
\cite{S88,G88,S95}), the most suitable action for the
description of a tensor multiplet interacting with a superstring
is provided by the Weyl invariant de Wit--Ro$\check{c}$ek action
\cite{dWR}, Eq. (\ref{ITMS}),
\begin{eqnarray}\label{Phi-ac}
& S_{\Phi}  = s
\int d^8Z \, E\, {\Phi \over 2} e^{{\Phi \over 2}} \; , \\ \nonumber
& ({\cal D}{\cal D}-\bar{R}) e^{{\Phi\over 2}}=0 \; , \quad
(\bar{{\cal D}}\bar{{\cal D}}-{R}) e^{{\Phi\over 2}}=0 \; .
\end{eqnarray}
The variation of  such an action with respect  to the superfield
$\Phi$ constrained by (\ref{Phi=lnL})  and with respect to
the supergravity multiplet is technically quite involved. However, the
following observation helps. The two--form satisfying the constraints
(\ref{H3}) is expressed essentially in terms of the tensor
multiplet superfield $e^{{\Phi\over 2}}$. Thus the problem of
varying the real linear multiplet is equivalent to the problem of
finding admissible variations of the two--form $B_2$ satisfying
the constraints (\ref{H3}). Such a task appears to be more
algorithmic. Moreover, for the variation of the interacting action
we will, anyway, need the form of the admissible variations of
$B_2$ superform, as its pull--back defines the Wess--Zumino term
of the Green--Schwarz superstring (\ref{GSac}).

\subsection{Varying the tensor multiplet.
Admissible variations of the two--form gauge superfield}

Clearly the constraints (\ref{H3}) make it impossible to consider the
variations of the two--form $B_2$ as independent. One rather has
to define ({\it cf.} Sec. IIA)
\begin{eqnarray}\label{vBdef}
\delta B_2 = {1\over 2} E^A\wedge E^B b_{BA}(\delta)
\end{eqnarray}
and find the expressions for  $b_{BA}(\delta)=  -(-1)^{BA} b_{AB}(\delta)$
from the conditions of conservation of the constraints (\ref{H3}).
Factoring out the gauge transformations $\delta^{gauge} B_2= d\alpha_1$,
one finds after tedious calculations
\begin{eqnarray}
\label{vBss}
& b_{\alpha\beta}(\delta)=0 \; , \quad b_{{\alpha}\dot{\beta}}(\delta)=0 \; ,
\quad b_{\dot{\alpha}\dot{\beta}}(\delta)=0 \; ,
\\ \label{vBsv}
& b_{\beta b}(\delta)= \; \sigma_{b\beta \dot{\beta}}
({\cal D}{\cal D}- \bar{R}) \delta \bar{\nu}^{\dot{\beta}} -
\qquad \nonumber \\ & \qquad - {i\over 2}
(\eta_{ab}\delta + \sigma_b\tilde{\sigma}_a)_\beta{}^{\gamma}
\nabla_{\gamma}e^{{\Phi \over 2}}\, \delta H^a \; ,
\\ \label{vBbsv}
& b_{\dot{\beta} b}(\delta)= - \sigma_{b\beta \dot{\beta}}
(\bar{{\cal D}}\bar{{\cal D}}- {R}) \delta {\nu}^{{\beta}} +
\qquad \nonumber \\ & \qquad +
{i\over 2} (\eta_{ab}\delta + \tilde{\sigma}_a\sigma_b)^{\dot{\gamma}}
{}_{\dot{\beta}}
\bar{\nabla}_{\dot\gamma}e^{{\Phi \over 2}}\, \delta H^a \; ,
\\ \label{vBvv}
& b_{a b}(\delta)  = -{i\over 4}
(\tilde{\sigma}_{[a}\sigma_{b]})^{\dot\beta}{}_{\dot\alpha}
\bar{{\cal D}}_{\dot\beta}
({\cal D}{\cal D} -\bar{R})
\delta \bar{\nu}^{\dot\alpha} - \qquad \nonumber \\
 & \qquad - {i\over 4}
(\sigma_{[a}\tilde{\sigma}_{b]})_{\alpha}{}^{\beta}
{{\cal D}}_{\beta} (\bar{{\cal D}}\bar{{\cal D}} -{R}) \delta {\nu}^{\alpha}
\; \nonumber \\
 & \qquad + {1\over 2}e^{{\Phi \over 2}}
\tilde{\sigma}_{[a}^{\dot{\beta}\beta} [{\cal D}_{\beta},
\bar{{\cal D}}_{\dot{\beta}}]\delta H_{b]} - {i\over 2}
\epsilon_{abcd} {\cal D}^c e^{{\Phi \over 2}} \, \delta H^d
\end{eqnarray}

As the components of the superfield strength of the two--form $B_2$
are expressed through the dilaton superfield, it should not be a
surprise that the preservation of the constraints (\ref{H3})
defines as well the variation of the dilaton superfield, 
\begin{eqnarray}
\label{vPhi}
\delta e^{{\Phi \over 2}} & = {i\over 2} \bar{{\cal D}}_{\dot\alpha}
({\cal D}{\cal D} -\bar{R}) \delta \bar{\nu}^{\dot\alpha} -
{i\over 2} {\cal D}_{\alpha}
(\bar{{\cal D}}\bar{{\cal D}} -{R}) \delta {\nu}^{\alpha} - \quad \nonumber \\
& - {1\over 4}
\tilde{\sigma}_a^{\dot{\beta}\beta} [{\cal D}_{\beta},
\bar{{\cal D}}_{\dot{\beta}}]e^{{\Phi \over 2}}\;
\delta H^a \, - 2 e^{{\Phi \over 2}}
(\Lambda (\delta)+ \bar{\Lambda}(\delta))\; , \qquad 
\end{eqnarray}
where $(\Lambda (\delta)+ \bar{\Lambda}(\delta))$ is defined in
Eq. (\ref{Lb+cc}).

\section{Interacting action and superfield equations of motion }
\setcounter{equation}0

Now that we have found all the necessary basic variations, we may turn
to varying the coupled  action
\begin{eqnarray}\label{Sint}
& S & = \int d^8 Z \, sdet(E_M{}^A)\,  (1 + s {\Phi \over 2}
e^{{\Phi  \over 2}}) +  S_{sstr} \; , \qquad {}\;
\\ \label{1GSac}
& S_{sstr} & =
\int_{W^2} [{1 \over 4}
* \hat{E}_a \wedge \hat{E}^{a} e^{{\hat{\Phi}\over 2}}
- \hat{B}_2] \; , \quad
\end{eqnarray}
to derive the equations of motion.

\subsection{Superstring equations}

Clearly, the superstring equations of motion for the interacting
system keep the same 
form as the superstring equations in the superspace background of
the superfield supergravity and tensor multiplet,
 \begin{eqnarray}
\label{EqSf} & (*\hat{E}^a - \hat{E}^a) \wedge
(\sigma_{a\alpha\dot{\alpha}} \hat{\bar{E}}{}^{\dot{\alpha}} -
{i\over 8} \hat{E}^b (\sigma_a\tilde{\sigma}_b)_{\alpha}{}^{\beta}
\hat{\nabla}_\beta \hat{\Phi}) =0\, , \nonumber  \\ & {}
\hspace{2.0cm} and\; \qquad  c.c.\; ,
\\
\label{EqSb} & {\cal D}(e^{{\hat{\Phi}\over 2}}*\hat{E}^a) +
\hat{E}^c \wedge  \hat{E}^b \, \hat{H}_{abc} - \hspace{3.2cm}
\nonumber \\ & {} \quad - {1\over 2}  *\hat{E}_b \wedge  \hat{E}^b
\, \hat{\nabla}_a e^{{\hat{\Phi}\over 2}} + \;  \hat{E}^c \wedge
\hat{E}^{\alpha}
(\sigma_{[a}\tilde{\sigma}_{b]})_{\alpha}{}^{\beta}
\hat{\nabla}_\beta e^{{\hat{\Phi}\over 2}} - \; \hspace{-0.2cm}
\nonumber \\ & {} \hspace{2.5cm}  - 2i \hat{E}^{\alpha} \wedge
\hat{\bar{E}}{}^{\dot{\alpha}} \sigma_{a\alpha\dot{\alpha}}
e^{{\hat{\Phi}\over 2}} =0 \; . \quad
\end{eqnarray}

\subsection{Superfield equations for tensor multiplet}

The equations of motion for the tensor multiplet appear as a result of
the $\delta \nu^\alpha$ and $\delta \bar{\nu}^{\dot{\alpha}}$ variations of the
dilaton superfield, Eq. (\ref{vPhi}), and the 2--superform $B_2$,
Eqs.(\ref{vBss})--(\ref{vBvv}). They are
\begin{eqnarray}\label{vnua}
 s (\bar{{\cal D}}\bar{{\cal D}}-R)&
(e^{-{\Phi\over 2}}{\cal D}_\alpha e^{{\Phi\over 2}})
 = - (\bar{{\cal D}}\bar{{\cal D}}-R) {\cal D}_\alpha K_a{}^a +
\nonumber \\ &
+ 4i \sigma_{b\alpha\dot{\beta}}
(\bar{{\cal D}}\bar{{\cal D}}-R) W^{b\dot{\beta}} -
\nonumber \\ &
- {1\over 2} (\sigma_{[a}\tilde{\sigma}_{b]})_\alpha{}^{\beta}
(\bar{{\cal D}}\bar{{\cal D}}-R) {\cal D}_{\beta}W^{ab}\; ,
\\
\label{vbnuda}
s({\cal D}{\cal D}-\bar{R})&
(e^{-{\Phi\over 2}}\bar{{\cal D}}_{\dot{\alpha}} e^{{\Phi\over 2}})
 = - ({\cal D}{\cal D}-\bar{R}) \bar{{\cal D}}_{\dot{\alpha}} K_a{}^a +
\nonumber \\ &
+4i \sigma_{b\beta\dot{\alpha}}
({\cal D}{\cal D}-\bar{R})  W^{b{\beta}} +
\nonumber
\\ &
+ {1\over 2} (\tilde{\sigma}_{[a}{\sigma}_{b})^{\dot{\beta}}{}_{\dot{\alpha}}
({\cal D}{\cal D}-\bar{R}) \bar{{\cal D}}_{\dot\beta} W^{ab}\; ,
\end{eqnarray}
where
\begin{eqnarray}\label{W}
& W^{BA} & :=  {1\over 2} \int_{W^2} {1\over \hat{E}} \,
\hat{E}^B \wedge \hat{E}^A \, \delta^8(Z-\hat{Z})\; \qquad
\end{eqnarray}
are current prepotentials which appear naturally in any variation of the
Wess--Zumino term of the
superstring action. In the same manner, any variation of the Nambu--Goto terms
of the superstring action will be expressed through the current prepotential
\begin{eqnarray}
\label{K}
& K_a^{B} & := {1\over 4}
\int_{W^2} {e^{{\hat{\Phi}\over 2}}\over \hat{E}} \, 
\ast\hat{E}_a \wedge  \hat{E}^B\,
\delta^8(Z-\hat{Z})\;  \qquad
\end{eqnarray}
({\it cf.} with the superparticle current prepotentials in \cite{BAIL2}).

\subsection{Superfield supergravity equations}

Now let us turn to the supergravity equations for the coupled system,
which appear
as a result of the $\delta {\cal H}^a$, $\delta {\cal U}$,
$\delta \bar{{\cal U}}$ variations
(see Sec.II for free supergravity).

The first observation is that, in accordance with
(\ref{vPhi})  and (\ref{varEa}), the variation of the Nambu--Goto terms
of the superstring action with respect to supergravity superfields does not
contain an input from $\Lambda(\delta)$, $\bar{\Lambda}(\delta)$,
defined in Eq. (\ref{Lb}), (\ref{Lb+cc}),
\begin{eqnarray}\label{vU-str}
& \delta_{{\cal U}} S_{sstr}  =
\int_{W^2} [ {1 \over 2}
* \hat{E}_a \wedge \delta_{{\cal U}} \hat{E}^{a} e^{{\hat{\Phi}\over 2}} +
{1 \over 4}
* \hat{E}_a \wedge \hat{E}^{a} \delta_{{\cal U}} e^{{\hat{\Phi}\over 2}}] =
\nonumber \hspace{-1cm} \\ & \; = {1 \over 2} \int_{W^2}
* \hat{E}_a \wedge \hat{E}^{a} e^{{\hat{\Phi}\over 2}}
({\cal D}{\cal D}-\bar{R})\delta {\cal U}
\nonumber \hspace{-1cm} \\ & \; - {1 \over 4} \int_{W^2}
* \hat{E}_a \wedge  \hat{E}^{a} e^{{\hat{\Phi}\over 2}}
\, 2 ({\cal D}{\cal D}-\bar{R})\delta {\cal U} = 0 \; .\hspace{-1cm}
\end{eqnarray}
Clearly, no such inputs come from the variations of the
pull--back $\hat{B}_2$ of the two--form $B_2$, as Eqs.
(\ref{vBss})--(\ref{vBvv}) does not contain $\Lambda(\delta)$,
$\bar{\Lambda}(\delta)$ at all. As the chiral variations $({\cal
D}{\cal D}-\bar{R})\delta {\cal U}$ and c.c. are involved in the
variations of the supervielbein only inside the $\Lambda(\delta)$,
$\bar{\Lambda}(\delta)$ combinations, this means that the
equations ${\delta S \over \delta {\cal U}}=0$ and ${\delta S
\over \delta \bar{{\cal U}}}=0$ do not possess an input from the
superbrane source (as was also the case with the supergravity---superparticle 
system, see \cite{BAIL2}). These equations can acquire an input from the
action  of the real linear multiplet, which in general, has the
form
 $s\int d^8Z E f({\Phi \over 2})$ with an arbitrary function $f$. However,
one can check that an input in the $\delta {\cal U}$ variation of
the improved kinetic term,   $s\int d^8Z E {\Phi \over 2} e^{{\Phi
\over 2}}$, also vanishes
\begin{eqnarray}\label{vU-SPhi}
& \delta_{{\cal U}} \int d^8Z E {\Phi \over 2} e^{{\Phi \over 2}}
= 0 \; .
\end{eqnarray}
Thus for the coupled action (\ref{Sint}) one finds that the
chiral superfield equation  ${\delta S \over \delta {\cal U}}=0$ remains the
same as in the case of `free' supergravity,
\begin{eqnarray}
\label{eqmbR}
{\delta S\over \delta {{\cal U}}}=0 \quad \Rightarrow
\quad \bar{R}=0 \; ,
\\
\label{eqmR}
{\delta S\over \delta \bar{{\cal U}}}=0 \quad \Rightarrow \quad
 R=0 \; .
\end{eqnarray}

Thus, in this case, as in the case of supergravity--massless
superparticle \cite{BAIL2}, only the vector superfield
supergravity equation ${\delta S\over \delta H^a}=0$ acquires a
source term from the superstring. However, in the coupled system
under consideration, these equations are more complicated due to
the supergravity interaction with the tensor multiplet,
\begin{eqnarray}
\label{eqmG}
{ \delta S \over \delta H^a }=0 & & \Rightarrow 
\qquad \nonumber \\
  G_a \, (1- s e^{ {\Phi\over 2} } )  &=&  {\cal J}_a \, + 
 \hspace{2cm}    \nonumber \\  &+& s
(5+ 3 {\Phi\over 2})  \tilde{\sigma}_a^{\dot{\beta}\beta}
[{\cal D}_{\beta}, \bar{{\cal D}}_{\dot\beta}] e^{{\Phi\over 2}} +
\qquad \nonumber \\
&
+& 3s \tilde{\sigma}_a^{\dot{\beta}\beta}  e^{-{\Phi\over 2}}\,
{\cal D}_{\beta}e^{{\Phi\over 2}}\, \bar{{\cal D}}_{\dot\beta}
e^{{\Phi\over 2}}
 \; .
\end{eqnarray}
The superstring current potential
\begin{eqnarray}
\label{Ja=def}
{\cal J}_a= - 6
{\delta S_{sstr}\over \delta H^a}
\end{eqnarray}
 entering the
{\it r.h.s} of Eq. (\ref{eqmG}), can be expressed
through
{\sl two types} of current prepotentials, Eqs. (\ref{K}) and (\ref{W}),
as follows
\begin{eqnarray}
\label{J=DDK}
- {1\over 6} {\cal J}_a = & 2i {\cal D}_{\beta} K_a^{\beta} - 2i
\bar{{\cal D}}_{\dot\beta} K_a^{\dot\beta} -
\tilde{\sigma}_b^{\dot{\beta}\beta}
[{\cal D}_{\beta}, \bar{{\cal D}}_{\dot\beta}] K_a{}^b -
\nonumber \\
& -{1\over 4}  K_b{}^b \tilde{\sigma}_a^{\dot{\beta}\beta}
[{\cal D}_{\beta}, \bar{{\cal D}}_{\dot\beta}]e^{{\Phi\over 2}} + \qquad
\nonumber \\
& + i W^{b\alpha} (\eta_{ba} \delta +
(\sigma_{b}\tilde{\sigma}_{a})_{\alpha}{}^{\beta}
\nabla_{\beta}  e^{{\Phi \over 2}} -
\nonumber \\
&
- i  W^{b\dot\alpha}
(\eta_{ba} \delta + (\tilde{\sigma}_{a}\sigma_{b})^{\dot\beta}{}_{\dot\alpha}
\bar{\nabla}_{\dot\beta}  e^{{\Phi \over 2}} +
\nonumber \\
+ {1\over 2} \tilde{\sigma}^{b\dot{\beta}\beta}&\!\!\!
[{\cal D}_{\beta}, \bar{{\cal D}}_{\dot\beta}] (e^{{\Phi\over 2}}W_{ab})
+ {i\over 2} \epsilon_{abcd} W^{bc} \nabla^d e^{{\Phi\over 2}} \; . \quad 
\end{eqnarray}
The first line of the {\it r.h.s} of Eq. (\ref{J=DDK}) has exactly
the same form as the expression for the current through the
superparticle current potential in the
supergravity---superparticle coupled system \cite{BAIL2}; the
second line contains the trace $K_b{}^b$  of the bosonic current
potential (\ref{K}) which vanishes in the superparticle case but
is nonzero for the superstring; the remaining part of the  {\it
r.h.s} of Eq. (\ref{J=DDK})  contains the current prepotentials
(\ref{W}) which come from the variation of the superstring
Wess--Zumino term.

Note that on the shell of Eqs. (\ref{eqmbR}), (\ref{eqmR}),
$R=0=\bar{R}$, the equations for the tensor multiplet
(\ref{vnua}), (\ref{vbnuda}) simplify to
\begin{eqnarray}\label{DDL}
 \bar{{\cal D}}\bar{{\cal D}}
[se^{-{\Phi\over 2}}{\cal D}_\alpha e^{{\Phi\over 2}}
& + {{\cal D}}_{{\alpha}} K_a{}^a
- 4i \sigma_{b\alpha\dot{\beta}} W^{b\dot{\beta}} +
\nonumber \\ &
+ {1\over 2} (\sigma_{[a}\tilde{\sigma}_{b]})_\alpha{}^{\beta}
{\cal D}_{\beta}W^{ab}]
=0\; ,
\\
\label{bDbDL}
{\cal D}{\cal D}[se^{-{\Phi\over 2}}\bar{{\cal D}}_{\dot{\alpha}}
e^{{\Phi\over 2}}
& +  \bar{{\cal D}}_{\dot{\alpha}} K_a{}^a
- 4i \sigma_{b\beta\dot{\alpha}} W^{b{\beta}} -
\nonumber
\\ &
- {1\over 2} (\tilde{\sigma}_{[a}{\sigma}_{b})^{\dot{\beta}}{}_{\dot{\alpha}}
\bar{{\cal D}}_{\dot\beta} W^{ab}]=0 \; .
\end{eqnarray}

\subsection{Superfield generalization of
the Einstein and Rarita--Schwinger equations
with sources}

In the minimal (off-shell) supergravity the superfield generalization of the
Ricci tensor and of the Rarita--Schwinger spin--tensor  are expressed
through  the vector and chiral scalar superfields as follows (see
\cite{BW} as well as \cite{BAIL2} and refs. therein)
\begin{eqnarray}
\label{Rici=off}
R_{bc}{}^{ac}& = {1\over 32} ({{\cal D}}^{{\beta}}
\bar{{\cal D}}^{(\dot{\alpha}|} G^{{\alpha}|\dot{\beta})} -
\bar{{\cal D}}^{\dot{\beta}} {{\cal D}}^{({\beta}}G^{{\alpha})\dot{\alpha}})
\sigma^a_{\alpha\dot{\alpha}}\sigma_{b\beta\dot{\beta}}
- \hspace{-1cm}
\nonumber \\ & - {3\over 64} (\bar{{\cal D}}\bar{{\cal D}}\bar{R}
+ {{\cal D}}{{\cal D}}{R}- 4 R\bar{R})\delta_b^a\; ,
\end{eqnarray}
\begin{eqnarray}\label{RS=off}
\Psi^a_{\dot{\alpha}} & :=
\epsilon^{abcd}T_{bc}{}^{\alpha}\sigma_{d\alpha\dot{\alpha}}= \hspace{2cm}
\nonumber \\
& ={i\over 8} \tilde{\sigma}^{a\dot{\beta}\beta} \bar{{\cal D}}_{(\dot{\beta}|}
G_{\beta|\dot{\alpha})} +
{3i\over 8} {\sigma}^a_{\beta \dot{\alpha}} {\cal D}^{\beta}R \; .
\end{eqnarray}

On the mass shell of the interacting system, taking into
account the superfield equations
of motion (\ref{eqmR}), (\ref{eqmbR}),
one finds that
the scalar curvature vanishes
\begin{eqnarray}
\label{Rsc=0}
R_{ab}{}^{ab}=0 \; ,
\end{eqnarray}
the Ricci tensor (\ref{Rici=off}) and the Rarita--Schwinger
spin--tensor (\ref{RS=off}) simplify. Then, to obtain the
superfield generalization of the Einstein and Rarita--Schwinger
equations for the interacting system one substitutes in Eqs.
(\ref{Rici=off}), (\ref{RS=off}) with $R=0=\bar{R}$ the formal
solution
\begin{eqnarray}
\label{eqmGs}
  G_a & = &
{1\over 1- s e^{ {\Phi\over 2} } } \; {\cal J}_a  + {\cal G}_a (\Phi)
\qquad
\end{eqnarray}
of the superfield equation (\ref{eqmG}). In (\ref{eqmGs}) ${\cal
G}_a (\Phi)$ denotes the on--shell `value' of the $G_a$ superfield
in the system of supergravity interacting with a dynamical tensor
multiplet, {\it i.e.} in the absence of the superstring,
\begin{eqnarray}
\label{calGs}
{\cal G}_a (\Phi) & := &
s {5+ 3 {\Phi\over 2} \over 1- s e^{ {\Phi\over 2} } }
\tilde{\sigma}_a^{\dot{\beta}\beta}
[{\cal D}_{\beta}, \bar{{\cal D}}_{\dot\beta}] e^{{\Phi\over 2}} +
\qquad \nonumber \\
&+&  {3s\over 1- s e^{ {\Phi\over 2} } }
\tilde{\sigma}_a^{\dot{\beta}\beta} e^{-{\Phi\over 2}}\,
{\cal D}_{\beta}e^{{\Phi\over 2}}\, \bar{{\cal D}}_{\dot\beta}
e^{{\Phi\over 2}}
 \; .
\end{eqnarray}

Thus the superfield generalizations of the Rarita--Schwinger and
the Einstein equations in the {\sl supergravity--tensor
multiplet---superstring} interacting system read
\begin{eqnarray}\label{RS=on}
\Psi^a_{\dot{\alpha}} & :=&
\epsilon^{abcd}T_{bc}{}^{\alpha}\sigma_{d\alpha\dot{\alpha}}=
{i\over 8} \tilde{\sigma}^{a\dot{\beta}\beta}
\bar{{\cal D}}_{(\dot{\beta}|}
{\cal G}_{\beta|\dot{\alpha})}(\Phi) +
\nonumber \\
& +&
{i\over 8} \tilde{\sigma}^{a\dot{\beta}\beta}
\bar{{\cal D}}_{(\dot{\beta}|} \left(
{{\cal J}_{\beta|\dot{\alpha})}/( 1-se^{{\Phi\over 2}})}
\right)
\end{eqnarray}
and
\begin{eqnarray}
\label{Rici=on}
R_{bc}{}^{ac}& = &
{1\over 32} \left({{\cal D}}^{{\beta}}
\bar{{\cal D}}^{(\dot{\alpha}|} {\cal G}^{{\alpha}|\dot{\beta})}(\Phi)
- \right. \nonumber \\ & & \quad - \left.
\bar{{\cal D}}^{\dot{\beta}} {{\cal D}}^{({\beta}}
{\cal G}^{{\alpha})\dot{\alpha}} (\Phi)\right)
\sigma^a_{\alpha\dot{\alpha}}\sigma_{b\beta\dot{\beta}}
+ \quad
\nonumber \\ &
+&  {1\over 32} \left({\cal D}^{{\beta}}
\bar{{\cal D}}^{(\dot{\alpha}|}
[{\cal J}^{{\alpha}|\dot{\beta})} /(1-se^{{\Phi\over 2}})]
- \right.
\nonumber \\
&& \quad -  \left.
\bar{{\cal D}}^{\dot{\beta}} {{\cal D}}^{({\beta}}
[{\cal J}^{{\alpha})\dot{\alpha}} /(1-se^{{\Phi\over 2}})]\right)
\sigma^a_{\alpha\dot{\alpha}}\sigma_{b\beta\dot{\beta}} \; . 
\qquad {} \quad  
\end{eqnarray}

The spacetime Einstein and Rarita--Schwinger  equations can be obtained
as the leading ($\theta=0$) components of the superfield equations
 (\ref{Rici=on}), (\ref{RS=on}) in the Wess-Zumino gauge (see
\cite{WZ78,BW,BAIL2}, refs. therein
and also Sec. VA below).
One should note that
\begin{eqnarray}\label{TbbfWZ}
T_{ab}{}^{\alpha}\vert_{\theta =0} & = 2 e_a^\mu e_b^\nu
{\cal D}_{[\mu}\psi_{\nu]}^{\alpha}(x) - \hspace{2.5cm} \nonumber \\
& - {i\over 4} (\psi_{[a}\sigma_{b]})_{\dot{\beta}}
G^{\alpha\dot{\beta}}\vert_{\theta =0}
 - {i\over 4} (\tilde{\sigma}_{[a}\bar{\psi}_{b]})^{\alpha}
R\vert_{\theta =0} \; \qquad 
\end{eqnarray}
differs from the standard definition of the gravitino field strength,
${\cal D}_{[\mu}\psi_{\nu]}^{\alpha}(x)=
\partial_{[\mu} \psi_{\nu]}^{\alpha}(x) -
\psi_{[\nu}^{\beta}(x) \, w_{\mu]\beta}{}^{\alpha}\vert_{\theta =0}$
by the leading components of the main superfields $G^a\vert_{\theta =0}$ and
$R\vert_{\theta =0}$ only. In our case on the mass shell
$R\vert_{\theta =0}=0$ (see (\ref{eqmR}))
and the $G^a$ superfield is determined by Eqs.
(\ref{eqmGs}),
(\ref{calGs}). Thus
\begin{eqnarray}\label{TWZ+th}
T_{ab}{}^{\alpha}\vert_{\theta =0} & = 2 e_a^\mu e_b^\nu
{\cal D}_{[\mu}\psi_{\nu]}^{\alpha}(x) -
{i\over 4} (\psi_{[a}\sigma_{b]})_{\dot{\beta}}
{\cal G}^{\alpha\dot{\beta}}(\Phi)\vert_{\theta =0} -
\hspace{1cm} \hspace{-1.5cm}
\nonumber \\ &
- {i\over 4(1-se^{\phi(x)\over 2})}
(\psi_{[a}\sigma_{b]})_{\dot{\beta}}
{\cal J}^{\alpha\dot{\beta}}(\Phi)\vert_{\theta =0}
\; , \qquad 
\end{eqnarray}
where $\phi(x)=\Phi\vert_{\theta =0}$.

\subsection{Superfield generalization of the 
Kalb--Ramond gauge field  equations
with source for the interacting system}

Taking the vector covariant derivative of the expression
(\ref{Habc}) for $H_{abc}$, one finds the off-shell expression for
the {\it l.h.s} of the 2-superform gauge field equation,
\begin{eqnarray}\label{D*H0}
{\cal D}^c H_{abc} &=& {1\over 32}
({\sigma}_{[a}\tilde{\sigma}_{b]})^{\alpha\beta}
{\cal D}_{(\alpha } {\bar{\cal D}}{\bar{\cal D}}
{\cal D}_{\beta )} e^{{\Phi\over 2}} \, + \, c.c. \, - \nonumber \\
 &-& {1\over 2} [ {\cal D}_a\, , \, {\cal D}_b] e^{{\Phi\over 2}}
 + {1\over 2} G_{[a}{\cal D}_{b]} e^{{\Phi\over 2}}
 + {1\over 2} H_{abc} G^c +
\nonumber \\
&+& {5\over 32}  \epsilon_{abcd} {\cal D}^c (e^{{\Phi\over 2}} G^d) -
\nonumber \\
&-&
{i\over 64}  ({\sigma}_{[a}\tilde{\sigma}_{b]})^{\alpha\beta}
W_{\alpha\beta\gamma}   {\cal D}^{\gamma}   e^{{\Phi\over 2}} \, + c.c. \; .
\end{eqnarray}
To arrive from (\ref{D*H0}) at the (superfield generalization) of
the antisymmetric tensor gauge field equations (Kalb--Ramond equations), 
we shall
substitute the expression for ${\cal D}_{(\alpha } {\bar{\cal
D}}{\bar{\cal D}} {\cal D}_{\beta )} e^{{\Phi\over 2}}$ which
follows from acting by the spinor covariant derivative ${\cal
D}_{\beta }$ on the superfield equations of motion (\ref{DDL}) for
the   tensor multiplet and, then, substitute (\ref{eqmGs}) for
$G^a$. The equations thus obtained have quite a complicated form.
Writing explicitly only the terms with the maximal number of the
spinor covariant  derivatives acting on the current prepotentials
(see below for a special r\^ole of such terms) one gets
\begin{eqnarray}\label{D*H}
s{\cal D}^c H_{abc} &=& - {1\over 32} e^{-{\Phi\over 2}}
({\sigma}_{[a}\tilde{\sigma}_{b]})^{\alpha\beta} {\cal
D}_{(\alpha} {\bar{\cal D}}{\bar{\cal D}}
{\cal D}_{\beta )} K_a{}^a\, + \nonumber \\
 &+& {i\over 8} e^{-{\Phi\over 2}}
 ({\sigma}_{[a}\tilde{\sigma}_{b]}{\sigma}_{c})_{\alpha\dot{\alpha}}
{\cal D}^{\alpha}{\bar{\cal D}}{\bar{\cal D}}W^{\dot{\alpha}c} \,
 - \nonumber \\
&-& {1\over 64}e^{-{\Phi\over 2}}
({\sigma}_{[a}\tilde{\sigma}_{b]}{\sigma}_{[c}
\tilde{\sigma}_{d]})_{\alpha}{}^{\beta} {\cal D}_{\beta}{\bar{\cal
D}}{\bar{\cal D}}{\cal D}^{\alpha} W^{cd} +\, \nonumber \\
&& \qquad + \; c.c.\, + \ldots \; .
\end{eqnarray} 
A further study of the complete form of the superfield equations for the
 tensor multiplet and for the Kalb--Ramon gauge field entering that multiplet
will be the subject of a separate paper.

Below we will show that the knowledge of the general form of the
tensor multiplet superfield equations with the source, Eq.
(\ref{DDL}), and of its relation to the gauge field equation
(through Eq.  (\ref{D*H})) already allow one  to make interesting
conclusions that provide a shortcut in the study of the
interacting system.

\section{Gauge equivalent description of the superfield
interacting system} \setcounter{equation}0

\subsection{Superdiffeomorphism symmetry gauge fixing}

The interacting action (\ref{Sint}) is manifestly invariant under
the local Lorentz symmetry and under
superdiffeomorphisms
\begin{eqnarray}
\label{sdZ} Z^{\prime M}&=& Z^M + b^M(Z)\; : \quad \begin{cases}
x^{\prime \mu}= x^\mu + b^\mu (x, \theta )\, , \\ \theta^{\prime
\breve{\alpha}}=  \theta^{\breve{\alpha}} +
\varepsilon^{\breve{\alpha}} (x, \theta) \; , \end{cases}\,
\\ \label{sdiffF} &&
E^{\prime A}(Z^\prime) =
E^{A}(Z), \quad
\Phi^\prime(Z^\prime)= \Phi(Z)
 \qquad
\nonumber \\
&& \qquad
w^{\prime ab}(Z^\prime)=
w^{ab}(Z)\; ,  \quad etc.\, , \qquad
\end{eqnarray}
which act on the superstring variables, coordinate functions
$\hat{Z}^{M}=\hat{Z}^{M}(\xi)\equiv (\hat{x}^{\mu}(\xi)\, , \,
\hat{\theta}^{\breve{\alpha}}(\xi))$, by the pull--back of the
transformations  (\ref{sdZ}),
 \begin{eqnarray}
\label{sdhZ} & \hat{Z}^{\prime M}= \hat{Z}^M + b^M(\hat{Z})\; : \;
\begin{cases} \hat{x}^{\prime \mu}(\xi)= \hat{x}^\mu + b^\mu (\hat{x} ,
\hat{\theta})\, , \cr
 \hat{\theta}^{\prime \breve{\alpha}}(\xi) =
\hat{\theta}^{\breve{\alpha}} + \varepsilon^{\breve{\alpha}}
(\hat{x}, \hat{\theta})\;  .  \end{cases}
\end{eqnarray}
The action  (\ref{Sint}) is also invariant under the
worldsheet reparametrizations and under the $\kappa$--symmetry
(\ref{kappaSGZ}), which act on  the coordinate functions only.

Thus, omitting the worldvolume reparametrization for simplicity,
the complete variation of the superstring coordinate function under the
local symmetries of the interacting action (\ref{Sint}) is given by
\begin{eqnarray}
\label{gshZ}
\delta  \hat{Z}^M(\xi) &=& b^M( \hat{Z}(\xi)) +
\delta_{\kappa} \hat{Z}^M(\xi) \; ,
\end{eqnarray}
where $\delta_{\kappa} \hat{Z}^M(\xi)$ is defined in (\ref{kappaSGZ}).

Now we observe \cite{BAIL2,BAIL3}
that the superdiffeomorphism symmetry can be used to fix the
`fermionic unitary gauge'
\begin{eqnarray}
\label{thGAUGE}
\hat{\theta}^{\check{\alpha}}(\xi) =0 \; \qquad \Leftrightarrow
\qquad \hat{Z}^M(\xi) = (\hat{x}^\mu(\xi)\, , \, 0\, ) \; .
\end{eqnarray}
Moreover, in the same manner as in  \cite{BAIL2}
one can show that {\it this gauge can be fixed simultaneously with
the Wess--Zumino gauge for supergravity} (see \cite{BZ85}, \cite{BAIL2}
and refs. therein),
 \begin{eqnarray}\label{WZgauge}
\theta^{\breve\alpha} E_{\breve\alpha}^{\, a}(x,\theta)=0\; , \qquad
\theta^{\breve\alpha} (E_{\breve\alpha}^{\, \underline{\beta}}(x,\theta)-
\delta_{\breve\alpha}^{\, \underline{\beta}})
=0\; , \\ \nonumber
\theta^{\breve\alpha} w_{\breve\alpha}^{ab}(x,\theta)=0\; ,
\end{eqnarray}
where, in particular, \cite{WZ78,BW}
\begin{eqnarray}\label{WZg}
E_\mu^{\; a}\vert_{_{\theta =0}} \propto e_\mu^{\; a}(x) \; , \qquad
E_\mu^{\; \underline{\alpha}}\vert_{_{\theta =0}} \propto
\psi_\mu^{\underline{\alpha}}\, ,
\\  \label{WZg=0}
E_{\breve{\beta}}^{\; a}\vert_{_{\theta =0}} =0  \; , \qquad
E_{\breve{\beta}}^{\; \underline{\alpha}}\vert_{_{\theta =0}}
= \delta_{\breve{\beta}}^{\; \underline{\alpha}}
\, ,
\\ \label{WZw}
w_\mu^{ab}\vert_{_{\theta =0}} \propto \omega_\mu^{ab}(x) \; .
\end{eqnarray}

This can be understood by observing that, although both 
gauges, Eqs. (\ref{WZgauge}) and (\ref{thGAUGE}), are fixed with
the use of the same superdiffeomorphism symmetry with the
parameter $b^M(Z)= b^M(x,\theta)$, the transformation rules of the
supergravity superfields involve only derivatives of
$b^M(x,\theta)$ (characteristic property of the gauge field
transformations), while the transformation rules of the coordinate
functions $\hat{Z}^M(\xi)$ (\ref{gshZ}) contain the additive
contribution of $b^M(\hat{Z}(\xi))=
b^M(\hat{x}(\xi),\hat{\theta}(\xi))$ (characteristic property of
the Goldstone field transformations, but for Goldstone fields
defined on a surface in superspace, see Sec. VIIB of \cite{BAIL2}
and \cite{BAIL3,BAILH}).

\subsection{Gauge fixed action}

Let us discuss what happens with the interacting action (\ref{Sint}),
(\ref{1GSac})
 in the
gauge (\ref{thGAUGE}), (\ref{WZgauge}). After integration over the Grassmann
coordinates, the Wess--Zumino supergravity action
(\ref{SGact}) becomes the standard supergravity action with the minimal set of
auxiliary fields \cite{auxN1},
\begin{eqnarray}\label{SGcom}
S_{SG} &=& \int d^4 x \tilde{d}^4\theta \; sdet(E_M^A)
\propto \nonumber \\
& \propto 
& S_{sg} =  \int d^4 x (e {\cal R} + \epsilon^{\mu\nu\rho\sigma}
\bar{\psi}_\mu\gamma_5\gamma_\nu {\cal D}_\rho \psi_\sigma)
\nonumber \\
& & + {\cal O} (g_a(x), r(x), \bar{r}(x))\; ,
\end{eqnarray}
where ${\cal R}= R_{\mu\nu}{}^{ab}(x) e_a^\mu (x) e_b^{\nu}(x)$ is
the scalar curvature of  spacetime, $e=det(e_\mu^a)$ and ${\cal O}
(g_a(x), r(x), \bar{r}(x))$ denotes  terms with auxiliary fields
\begin{eqnarray}\label{SGauxf}
G_a \vert_{_{\theta =0}} \propto g_a(x) \; , \quad
R\vert_{_{\theta =0}} \propto r(x) \; , \quad
\bar{R}\vert_{_{\theta =0}} \propto \bar{r}(x) \;  \;
\end{eqnarray}
which are not essential for the consideration below.
The improved tensor multiplet action (\ref{ITMS})
also entering (\ref{Sint}),
becomes, schematically, (see \cite{dWR})
\begin{eqnarray}\label{ITMcom}
S_{TM} & =  s\int d^8 Z \, E \, {\Phi(Z) \over 2} e^{{\Phi(Z)\over 2}}
\propto 
\hspace{2.8cm}
\nonumber \\
\propto   S_{tm} & =   \int d^4 x e [{1\over 2} e^{{\phi(x)\over 2}}
g^{\mu\nu}{\cal D}_\mu \phi {\cal D}_\nu \phi  +
{1\over 3!} {H}^{\mu\nu\rho} H_{\mu\nu\rho} +  \qquad \hspace{-1.0cm}
\nonumber \\
   & +
i e^{{\phi\over 2}}  ({\cal D}_a \chi^\alpha \sigma^a_{\alpha
\dot{\alpha}} \bar{\chi}{}^{\dot{\alpha}} + c.c.) + \ldots ]
\; , \qquad \hspace{-1.0cm}
\end{eqnarray}
where ${\cal D}_\mu$ denotes the spacetime covariant derivatives and
the component fields of the tensor multiplet are defined by
\begin{eqnarray}\label{ITMphi}
\Phi \vert_{_{\theta =0}}\propto \phi(x)\; , \qquad {\cal
D}_\alpha e^{\Phi\over 2}  \vert_{_{\theta =0}} \propto
\chi_\alpha (x)\; , \qquad
\end{eqnarray}
 and ({\it cf.} (\ref{H3}))
 \begin{eqnarray}\label{B(x)}
[{\cal D}_{\alpha}, \bar{{\cal D}}_{\dot{\alpha}}] e^{{\Phi\over
2}} \vert_{_{\theta =0}}  \propto \sigma_{a\alpha\dot{\alpha}}
e_\mu^a(x) \epsilon^{\mu\nu\rho\sigma} H_{\nu\rho\sigma} \; ,
\end{eqnarray}
$H_{\mu\nu\rho}(x) = 3\partial_{[\mu} B_{\nu\rho]}(x)$. In
(\ref{ITMcom}) we have written explicitly only the kinetic terms;
the  remaining ones may be extracted from the formulae in
\cite{dWR} and are not essential for what follows.

Finally, the superstring action (\ref{1GSac}) in the gauge (\ref{thGAUGE})
reduces to the {\it bosonic string action}
\begin{eqnarray}\label{BSac}
& S_{sstr}\vert_{\hat{\theta}=0} & \propto  S_{bstr} = 
\int_{W^2} [{1 \over 4}
* \hat{e}_a \wedge \hat{e}^{a} e^{{\hat{\phi}\over 2}}
- {B}_2(\hat{x})] \;  \qquad  \\
\nonumber
&& =  \int_{W^2} \left[{1 \over 2}
d^2 \xi \sqrt{|det(\hat{e}_m^a \hat{e}_{an})|}
e^{{{\phi}(\hat{x})\over 2}}
- {B}_2(\hat{x})\right] \; ,
\end{eqnarray}
where $$B_2(\hat{x}) = {1\over 2} d\hat{x}^\mu \wedge d\hat{x}^\nu
B_{\mu\nu}(\hat{x}) = d^2\xi \partial_\tau \hat{x}^\mu
\partial_\sigma \hat{x}^\nu B_{\mu\nu}(\hat{x})\; .$$

Thus the complete gauge fixed action for the interacting system reads
\begin{eqnarray}\label{SintGF}
S_{int\; GF}&=&
S_{sg} (e,\psi, g_a, r, \bar{r}) +
S_{tm} (\phi, \chi, B\; ; \; e, \psi, g_a) + \nonumber \\
&+& S_{bstr}(\hat{x}\; ; e,B)\; ,
\end{eqnarray}
with
$S_{sg}$, $S_{tm}$ and $S_{bstr}$ defined in (\ref{SGcom}), (\ref{ITMcom})
and (\ref{BSac}), respectively.

\subsection{Supersymmetry of the gauge fixed action}

Note that, although the gauge fixed action (\ref{SintGF}) includes the action 
for the purely bosonic string (\ref{BSac}), it possesses $1/2$ of the local 
supersymmetry characteristic for the supergravity action. Actually, the 
direct proof of this fact can be found in \cite{BdAI1}. 
Here we will show this in a different way which is based on  the observation 
that the symmetries of the gauge fixed action  (\ref{SintGF}) 
can be identified as a subset of the symmetries of the complete (superfield) 
action (\ref{Sint}) which preserve the gauge (\ref{thGAUGE}) 
and the Wess--Zumino gauge 
(see \cite{BAIL2} for the supergravity---superparticle interacting system). 

Firstly note that in the Wess--Zumino gauge 
(\ref{WZgauge})   
the index of the superspace Grassmann coordinate is identified 
with the Lorentz group spinor index. Indeed,  due to  
the second equation in (\ref{WZgauge}), 
\begin{eqnarray}\label{th-WZg}
\theta^{\underline{\beta}}\equiv (\theta^{{\beta}}, 
\bar{\theta}_{\dot{\beta}})
:= \theta^{\breve\alpha} E_{\breve\alpha}^{\, \underline{\beta}}(Z) =
\theta^{\breve\alpha} \delta_{\breve\alpha}^{\, \underline{\beta}}\; .
\end{eqnarray}
Clearly, the same is true for the  superstring 
spinorial Grassmann coordinate 
functions. Their  transformation rules can be read off Eq. (\ref{gshZ}), 
\begin{eqnarray}
\label{gshTh}
\delta  
\hat{\theta}^\alpha(\xi) &=& b^M ( \hat{x}, \hat{\theta}) 
E_M^\alpha(\hat{x}, \hat{\theta}) +
\delta_{\kappa} \hat{\theta}^\alpha(\xi) \; , \qquad 
\end{eqnarray}
where $\delta_{\kappa} \hat{\theta}^\alpha(\xi)$ reads 
(see Eqs. (\ref{kappaSGZ}) and (\ref{th-WZg})) 
\begin{eqnarray}
\label{kappathG}
\delta_{\kappa} \hat{\theta}^\alpha(\xi) &=& \bar{\kappa}^n_{\dot\alpha}
(\delta_n^m - \sqrt{|g|}
\epsilon_{nk}g^{km}) \hat{E}^a_m(\hat{x},\hat{\theta})  
\tilde{\sigma}_a^{\dot{\alpha}\alpha}
\; , \quad \nonumber \\ 
\delta_{\kappa} \hat{\bar{\theta}}{}^{\dot{\alpha}}(\xi) &=& 
(\delta_{\kappa} \hat{\theta}^\alpha(\xi) )^\ast \; . \qquad 
\end{eqnarray}
Clearly, the superdiffeomorphism parameter $b^M ({x}, {\theta})$ in 
(\ref{gshTh}) is to be restricted by the conditions of the preservation of 
the Wess--Zumino gauge. However (see \cite{WZ78}, \cite{BZ85}, 
\cite{BAIL2} and refs therein) the parameter 
$\epsilon^\alpha (x) = b^M ({x}, 0) E_M^\alpha({x}, 0)$  remains unrestricted 
and is identified with the parameter of the local supersymmetry of the 
component (spacetime) formulation of supergravity. 

Now, the preservation of the gauge (\ref{thGAUGE}), 
$\hat{\theta}^\alpha(\xi)= 0$, imposes the condition $\delta  
\hat{\theta}^\alpha(\xi)\vert_{\hat{\theta}^\alpha(\xi)= 0}=0$, {\it i.e.}  
\begin{eqnarray}
\label{gshTh1}
\epsilon^\alpha (\hat{x}) &=& - 
\delta_{\kappa}   
\hat{\theta}^\alpha(\xi)\vert_{\hat{\theta}^\alpha(\xi)= 0}= \nonumber  \\ 
&=& 
- \bar{\kappa}^n_{\dot\alpha}(\xi)
(\delta_n^m - \sqrt{|g|}
\epsilon_{nk}g^{km}) \hat{e}^a_m(\hat{x})  
\tilde{\sigma}_a^{\dot{\alpha}\alpha} \; , \qquad 
\end{eqnarray}
on the parameter of the local supersymmetry. This restriction appears only on 
the string worldsheet and expresses the pull--back $\epsilon^\alpha (\hat{x})$ 
of the supersymmetry parameter $\epsilon^\alpha ({x})$ 
through a worldsheet parameter $\bar{\kappa}^n_{\dot\alpha}(\xi)$ 
contracted (on both indices) with the expression 
$(\delta_n^m - \sqrt{|g|}
\epsilon_{nk}g^{km}) \hat{e}^a_m(\hat{x})  
\tilde{\sigma}_a^{\dot{\alpha}\alpha}$. The latter is the $\hat{\theta}=0$ 
`value' of the Green--Schwarz $\kappa$--symmetry `projector' and makes 
only one parameter included in $\bar{\kappa}^n_{\dot\alpha}(\xi)$ being 
involved {\sl effectively} in the expression. 

Thus on the worldsheet $W^2$ of a 
(dynamical) string the $4$ parameters of local 
supersymmetry of the free supergravity are reduced, by the condition of the 
invariance of the gauge fixed interacting action (\ref{SintGF}), 
to the $2$ effective parameters of the  
$\kappa$--symmetry--like transformations, while out of $W^2$ these $4$ 
parameters, $\epsilon^\alpha (x), \bar{\epsilon}^{\dot{\alpha}}(x)$,  
remain unrestricted. This can be characterized by stating  
the preservation of the $1/2$ of the local supersymmetry 
of the free supergravity by the gauge fixed interacting action  (\ref{SintGF}).

\subsection{On equations of motion following from the gauge fixed action}

An important observation is that the gauge fixed version of the
superstring action (\ref{BSac}) involves only the {\sl physical
bosonic fields} of the supergravity and tensor multiplet, the
graviton $e_\mu^a(x)$, the antisymmetric tensor $B_{\mu\nu}(x)$ and
the scalar $\phi(x)$. Neither auxiliary fields nor fermions appear
in the string action (\ref{BSac}). As a result, both the equations
for the auxiliary fields and for the fermions of the interacting
system (\ref{SintGF}) will keep formally the same form as in the
absence of superstring. In particular, this means that {\it in the
gauge (\ref{thGAUGE}), (\ref{WZgauge}) neither the
Rarita--Schwinger equations, nor the equations for the fermionic
field of the tensor multiplet will include a source term from the
superstring} (although they will be written with covariant
derivatives for the spin connections satisfying the sourceful
Einstein equations with an input from the superstring
energy--momentum tensor).

This is a manifestation of a counterpart of the
super--Higgs effect in dynamical supergravity interacting with
a superstring or superbrane object, which we will address in a separate
paper (see \cite{BAILH} for a spacetime counterpart of the Higgs effect
in general relativity interacting with material particles, strings and
p--branes).

Similar properties have already been found in the dynamical
$D=4$ $N=1$ supergravity interacting with a massless superparticle
\cite{BAIL2}. The present consideration generalizes it for the 
case of dynamical supergravity interacting with a simplest supersymmetric
extended object.

Below we present a simple check of the gauge equivalence described
above at the level of equations of motion.  Namely we will show
that the dynamical equations for fermions, which follow from the
superfield equations (\ref{eqmbR}), (\ref{eqmR}), (\ref{eqmG}),
(\ref{DDL}), are indeed sourceless in the gauge (\ref{thGAUGE}),
(\ref{WZgauge}).

\subsection{Superfield equations and the equations
for physical fields in the `fermionic unitary' gauge}

The superstring contributions to all the superfield equations of
motion, Eqs. (\ref{DDL}), (\ref{bDbDL}) and  (\ref{eqmG}) (or,
equivalently, (\ref{eqmGs}), (\ref{calGs})) with (\ref{J=DDK}),
come in the form of current prepotentials (\ref{K}), (\ref{W}), or
their derivatives. This is true as well for their  consequences
like Eqs. (\ref{RS=on}), (\ref{Rici=on}), (\ref{D*H}).

Both current prepotentials  (\ref{K}) and (\ref{W}) contain
the superspace delta function
\begin{eqnarray}
\label{delta}
\delta^8(Z-\hat{Z})=
\delta^4(x-\hat{x}) (\theta-\hat{\theta})^4
\end{eqnarray}
integrated over $W^2$ with the corresponding measure.
In the gauge (\ref{thGAUGE}), $\hat{\theta}^{\check{\alpha}}(\xi)=0$,
the delta function becomes proportional to the highest degree of the
Grassmann coordinate $\theta$
\begin{eqnarray}
\label{deltaGT}
\hat{\theta}^{\check{\alpha}}=0\; :  \qquad
\delta^8(Z-\hat{Z}) =
\delta^4(x-\hat{x}) \theta^4
\end{eqnarray}
As a result, in this gauge both superstring current prepotentials 
are proportional to the fourth degree of the Grassmann coordinate,
\begin{eqnarray}
\label{K=W=}
K_a^{B} \; \propto \; \theta^4 \; , \qquad
W^{AB} \; \propto \; {\theta}^4 \; .
\end{eqnarray}
Then a covariant derivative of any of the current prepotentials is
proportional
to  ${\theta}^3$,
\begin{eqnarray}
\label{DKW=}
{\cal D}_C K_a^{B} \; \propto \; \theta^3 \; , \qquad
{\cal D}_C W^{AB} \; \propto \; {\theta}^3 \; .
\end{eqnarray}
The action of two derivatives which is {\it not} reducible to one
derivative, ${\cal D}^2_{AB}= {\cal D}_A {\cal D}_B + (-1)^{AB}
{\cal D}_A {\cal D}_B$ but {\it not} $[ {\cal D}_A , {\cal D}_B\}
= - T_{AB}{}^C  {\cal D}_C + R_{AB}$ ({\it e.g.}, ${\cal
D}^2_{\alpha\dot{\beta}}\equiv [{\cal D}_{\alpha}, \bar{{\cal
D}}_{\dot{\beta}}]$, but not $\{ {\cal D}_{\alpha}, \bar{{\cal
D}}_{\dot{\beta}}\} = 2i \sigma^a_{\alpha\dot{\beta}} {{\cal
D}}_{a}$) may result in expressions
proportional to  ${\theta}^2$,
\begin{eqnarray}\label{D2KW=}
{\cal D}^2_{CD}
K_a^{B} \; \propto \; \theta^2 \; , \qquad
{\cal D}_{CD} W^{AB} \; \propto \; {\theta}^2 \; ,
\end{eqnarray}
{\it etc.},
\begin{eqnarray}\label{D3K=W=}
{\cal D}^3_{CDE} K_a^{B} \; \propto \; (\theta)^1 \; , \qquad
{\cal D}^3_{CDE} W^{AB} \; \propto \; ({\theta})^1 \; .
\end{eqnarray}
Only the action of four derivatives may produce a
$\propto \theta^0$ input, {\it i.e.} terms which have a nonvanishing
$\theta=0$ value,
\begin{eqnarray}\label{D4K=W=}
{\cal D}^4_{CDEF} K_a^{B} \propto \; \theta^0 \; , \quad
 {\cal D}^4_{CDEF}W^{AB} \propto \; {\theta}^0 \; .
\end{eqnarray}

This implies, in particular, that the current potential (\ref{J=DDK})
is proportional to the second power of the superspace Grassmann coordinate,
\begin{eqnarray}
\label{Jprop}
{\cal J}_a \; \propto \; \theta^2 \; .  \quad
\end{eqnarray}
Then
\begin{eqnarray}
\label{DJprop}
{\cal D}_A {\cal J}_a \; \propto \; \theta^1 \; ,   \quad
\end{eqnarray}
and only the second derivative of the current potential may produce
a term with a nonvanishing leading component
($\theta$--independent part),
\begin{eqnarray}
\label{DDJprop}
{\cal D}^2_{BC} {\cal J}_a \propto \theta^0 \;   \quad
\end{eqnarray}

The spacetime fermionic equations of motion of the interacting system
may be obtained as leading ($\theta=0$) components of the Eqs.
(\ref{RS=on}) and (\ref{DDL}).
Ignoring the inputs with a smaller number of derivatives applied to the 
current potential, one can write the leading components
of  Eq. (\ref{RS=on}) as
\begin{eqnarray}\label{RS=onS1}
(\Psi^a_{\dot{\alpha}} - \Xi^a_{\dot{\alpha}}(\Phi))\vert_{_{\theta =0}}
& \propto & {\cal D}_B {\cal J}_a \vert_{_{\theta =0}}
\end{eqnarray}
where $ \Xi^a_{\dot{\alpha}}(\Phi)$ denotes the tensor multiplet contribution 
to the gravitino equation, which is given by the first term in the
 {\it r.h.s.} of Eq. (\ref{RS=on}). Then Eq. (\ref{DJprop}) implies
that the {\it r.h.s.} of Eq. (\ref{RS=onS0}) vanishes in the gauge
(\ref{thGAUGE}), {\it i.e.} that in this gauge Eq. (\ref{RS=onS0}) reads
\begin{eqnarray}\label{RS=onS0}
\Psi^a_{\dot{\alpha}}\vert_{_{\theta =0}} =
\Xi^a_{\dot{\alpha}}(\Phi)\vert_{_{\theta =0}}
\end{eqnarray}
and does not contain an explicit input from the superstring.

In the same manner one finds that the dynamical equation for the fermionic
component of the tensor multiplet, given by the leading component of
(\ref{DDL})
\begin{eqnarray}\label{DDL0}
s \bar{{\cal D}}\bar{{\cal D}} [e^{-{\Phi\over 2}}
{\cal D}_\alpha e^{{\Phi\over 2}}]\vert_{_{\theta =0}}
&=& -
\bar{{\cal D}}\bar{{\cal D}}{{\cal D}}_{{\alpha}} K_a{}^a \vert_{_{\theta =0}}
 -
\nonumber \\
&-& {1\over 2} (\sigma_{[a}\tilde{\sigma}_{b]})_\alpha{}^{\beta}
\bar{{\cal D}}\bar{{\cal D}} {\cal D}_{\beta}W^{ab}] \vert_{_{\theta =0}}
\nonumber \\
&+& 4i \sigma_{b\alpha\dot{\beta}} \vert_{_{\theta =0}}
\bar{{\cal D}}\bar{{\cal D}} W^{b\dot{\beta}} \vert_{_{\theta =0}}
\; ,
\end{eqnarray}
implies, in the light of Eq. (\ref{D3K=W=}), that the superstring--produced
{\it r.h.s.} of this equation vanishes in the gauge (\ref{thGAUGE}),
\begin{eqnarray}\label{DDL0G}
s \bar{{\cal D}}\bar{{\cal D}} [e^{-{\Phi\over 2}}
{\cal D}_\alpha e^{{\Phi\over 2}}]\vert_{_{\theta =0}}
&=& 0 \; . \quad
\end{eqnarray}

As far as the bosonic superfield equations are concerned, they preserve the
nontrivial input from the superstring source in the gauge
(\ref{thGAUGE}). Indeed, the leading components of
Eqs. (\ref{Rici=on}), (\ref{D*H}) contain the second derivatives
of the current potential and fourth derivatives of the current prepotentials,
which remain nonvanishing in accordance with Eqs.
(\ref{DDJprop}) and (\ref{D3K=W=}).

The explicit form of the Einstein equation 
can be derived in a way close to the one used in \cite{BAIL2} for the 
{\sl supergravity---superparticle} interacting system (although the 
presence of both Nambu--Goto and Wess--Zumino terms 
in the superstring action, as well as of the tensor multiplet in the action 
of the interacting system, makes the Einstein equation 
of the {\sl supergravity---tensor multiplet---superstring} system 
a bit more complicated). As far as the Kalb--Ramon field equations are 
concerned, taking into account Eqs. (\ref{K=W=})--(\ref{D4K=W=}) and 
the conditions 
of the Wess--Zumino gauge (\ref{WZgauge}), (\ref{WZg}) 
(which implies, in particular, ${\cal D}_\alpha \bar{{\cal D}}\bar{{\cal D}}
{\cal D}^\beta (\theta)^4 \propto \delta_\alpha^\beta + 
{\cal O} (\theta)$) 
one finds that, in the gauge 
(\ref{thGAUGE}), the leading component of the superfield 
equation (\ref{D*H}) becomes 
\begin{eqnarray}\label{D*H1}
s{\cal D}^c H_{abc}\vert_{\theta =0} &=& 
 {1\over 16}e^{-{\phi\over 2}}
w^{cd} + \ldots \; ,
\end{eqnarray}
where 
\begin{eqnarray}\label{wba}
& w^{ba} & :=  {1\over 2} \int_{W^2} {1\over \hat{e}} \,
\hat{e}^b \wedge \hat{e}^a \, \delta^4(x-\hat{x})\; . \qquad
\end{eqnarray}
Hence, in the gauge (\ref{thGAUGE}) the superstring input on the 
{\it r.h.s.} of the Kalb--Ramond gauge field equation for the 
{\sl supergravity---tensor multiplet---superstring} interacting system 
is nonvanishing and, moreover, coincides with the input of the bosonic string.

Thus we have checked that the spacetime equations of motion for
the fermionic fields which follow from the complete superfield
action (\ref{Sint}) of the {\sl supergravity---tensor
multiplet---superstring} interacting system become sourceless in
the gauge (\ref{thGAUGE}). This is true both for the gravitino
equations and for the fermionic field of the tensor multiplet. At
the same time, the corresponding bosonic equations are clearly
sourceful in any gauge. 
This is a characteristic property of the
equations which follow from the gauge fixed action (\ref{SintGF}).

We should note that the gauge fixed fermionic equations are not
completely decoupled from the superstring. They are written in terms
of the spacetime covariant derivatives with (composed)
spin--connections satisfying the Einstein  equation with a 
source. The same is true for the equations derived directly from
the gauge fixed action (\ref{SintGF}).

\subsection{On possible application of the gauge equivalence}

Note that the gauge fixed description (\ref{SintGF}) of the
superfield interacting system (\ref{SintGF}) is complete in the
following sense. Along the line of \cite{BdAI1} one may check that
the gauge fixed action (\ref{SGcom}), (\ref{ITMcom}) reproduces
the gauge fixed version of all the dynamical equations which might
be derived from the complete superfield action, including the {\sl
fermionic equations for the bosonic string}. This is the
manifestation of the purely gauge (or Goldstone) nature of the
superstring coordinate functions $\hat{Z}^M(\xi)$ (not to be
confused with the supercoordinates $Z^M$; see \cite{BAILH} and
\cite{BAIL3} for further discussion).

A counterpart of the gauge fixed action (\ref{SintGF}) can be
written in any dimension, for any supergravity interacting with
any superbrane. Hence, using the above described gauge
equivalence  one may already proceed with studying 
the  $D=11$ supergravity interacting
with super--M2--branes and super--M5--branes, as well as the
$D=10$ type II supergravity interacting with super--Dp--branes
(in spite of the fact the $D=10,11$ superfield supergravity actions are not
known).

\section{Conclusions}
\setcounter{equation}0

 In this paper we studied the full
superfield Lagrangian description of the $D=4$ interacting system
of dynamical supergravity and a superstring described by the sum
$S=S_{SG}+ S_{TM} + S_{sstr}$ of the superfield supergravity
action $S_{SG}$ \cite{WZ78}, the Green--Schwarz superstring action
$S_{str}$ \cite{GS84} and a superfield action for the dynamical
tensor multiplet $S_{TM}$ \cite{dWR}.

The superfield theoretical system $S=S_{SG}+ S_{TM}$
was argued to be related to the low--energy limit of 
$D=4$ compactification of the 
heterotic superstring \cite{S88},\cite{G88} (see also \cite{G96}) 
and considered in \cite{S95,BS96}. So, our main interest here has been 
to analyze the influence of 
the superstring action $S_{sstr}$ on the dynamics of the interacting system, 
namely in the superstring--produced source terms both in the superfield 
and component form of the equations.

We have obtained the complete set of superfield equations with
sources provided by the superstring. In the supergravity sector we
found that the scalar superfield equation remains the same as for
free supergravity, while the vector superfield equation is
modified both by the interaction with the tensor multiplet and by
the source (current potential) coming from the superstring. The
current potential is constructed from the two types of current
{\it pre}potentials coming from the variation of the Nambu--Goto
and the Wess--Zumino terms of the superstring action,
respectively. The superfield equations for the tensor multiplet
are also modified by  inputs from the above mentioned current
prepotentials. The equations of motion for the supersting variables 
are the same  as in the {\sl background} of supergravity
interacting with dilaton and super--2--form superfields.

By analyzing the gauge symmetries and taking into account the properties 
of the  Wess--Zumino gauge (see \cite{BAIL2} and refs. therein)
we have shown that there exists a complete gauge equivalent
description of the `superfield' interacting system,  
 $S=S_{SG}+ S_{TM} + S_{sstr}$,  
given by the sum 
$S=S_{sg}+ S_{tm} + S_{bstr}$
of the 
spacetime {\it component} action for supergravity 
$S_{sg}$ (without auxiliary fields), the component action for 
the tensor multiplet $S_{tm}$ and  the action for a {\it bosonic} string
$S_{str}$
(which appears as the purely bosonic 
`limit' of the supersting action $S_{sstr}$).
We checked this gauge equivalence by studying the properties of
the gauge fixed version of the equations of motion derived from
the complete superfield action. Despite the quite complicated form
of the superfield generalizations of the Einstein and
Rarita--Schwinger equations, as well as of the Kalb--Ramond
equations and the equations for the fermionic fields of the tensor
multiplet, it turned out  quite easy to show that in the above
mentioned `fermionic unitary gauge' all the fermionic equations
are sourceless (although they include the covariant derivatives
with the spin connections obeying the sourceful Einstein
equations) while the bosonic equations, including the 
Einstein and Kalb--Ramond field equations, 
acquired a source from the supersting.

This extends the {\sl supergravity--massless superparticle} results of
\cite{BAIL2} to the case of dynamical interacting systems
including supergravity and an extended supersymmetric object and
stringly supports that the above mentioned gauge equivalence is not an
artifact of the simpler massless superparticle case, but rather is a general
property of the above mentioned 
interacting systems. As the
component actions for supergravity are known in all dimensions,
including 
$D=10, 11$ (the most interesting from an M--theoretic perspective), 
our results allows one to obtain and to study the
complete set of equations for dynamical supergravity interacting
with dynamical super--$p$--brane, at least in its gauge fixed
version. This promises to be a useful tool in a future search for
new solitonic solutions of higher dimensional supergravity
including the ones with nonvanishing fermionic fields.

An analysis of the superfield equations with sources
obtained in this paper, the search for their possible higher dimensional
generalization as well as the investigation of the $D=10, 11$
supergravity--superbrane interacting systems with the use of the gauge
equivalence of their complete superfield description with the description by
the sum of spacetime (component) action for supergravity and the action for
bosonic brane will be the subject of future work.

\bigskip

{\it Acknowledgments}. The authors are grateful to Jos\'e A. de
Azc\'arraga, D. Sorokin, J. Lukierski,  E. Ivanov, 
for useful conversations during different stages of this work 
and to S.J. Gates, P. Pasti, M. Tonin and W. Siegel for comments. 
This work has been partially supported by the research grant 
BFM2002-03681 from the
Spanish Ministerio de Ciencia y Tecnolog\'{\i}a and from EU FEDER
funds, by the grant N 383 of the Ukrainian State Fund for
Fundamental Research and by the INTAS Research Project N 2000-254.

\end{document}